\documentclass[journal]{IEEEtran}
\usepackage{amsmath,amsfonts}
\usepackage{algorithmic}
\usepackage{algorithm}
\usepackage{array}
\usepackage[caption=false,font=normalsize,labelfont=sf,textfont=sf]{subfig}
\usepackage{textcomp}
\usepackage{stfloats}
\usepackage{url}
\usepackage{verbatim}
\usepackage{graphicx}
\usepackage{cite}

\usepackage{graphicx}
\usepackage{booktabs}
\usepackage[multiple]{footmisc}
\usepackage{multirow}
\usepackage{array}
\usepackage{hyperref}
\usepackage{balance} 
\usepackage{pgfplots}
\usepackage{lipsum}
\usepackage{xcolor}
\usepackage{tikz}
\usepackage{soul}
\usepackage{enumitem}
\usepackage{bm}
\usepackage{tabularx}
\usepackage{amsmath}
\usepackage{kotex}

\def\BibTeX{{\rm B\kern-.05em{\sc i\kern-.025em b}\kern-.08em
    T\kern-.1667em\lower.7ex\hbox{E}\kern-.125emX}}

\pgfplotsset{
    name nodes near coords/.style={
        every node near coord/.append style={
            name=#1-\coordindex,
            alias=#1-last,
        },
    },
    name nodes near coords/.default=coordnode
}

\pgfplotsset{compat=1.5.1}
\newenvironment{customlegend}[1][]{%
  \begingroup
  \csname pgfplots@init@cleared@structures\endcsname
  \pgfplotsset{#1}%
}{%
  \csname pgfplots@createlegend\endcsname
  \endgroup
}%
\def\addlegendimage{\csname pgfplots@addlegendimage\endcsname}
\usetikzlibrary{patterns}
\usetikzlibrary{pgfplots.groupplots}
\usetikzlibrary{
  pgfplots.colorbrewer,
}

\definecolor{aa}{rgb}{0.2,0.7,0.310}
\definecolor{cc}{rgb}{1.0,0.49,0.0}
\definecolor{bb}{rgb}{0.514,0.325,0.831}

\begin{document}
\title{Trustworthiness-Driven Graph Convolutional Networks for Signed Network Embedding}

\author{Min-Jeong Kim\IEEEauthorrefmark{1},
        Yeon-Chang Lee\IEEEauthorrefmark{1},
        David Y. Kang,
        and~Sang-Wook Kim\IEEEauthorrefmark{3},~\IEEEmembership{Member,~IEEE}
\thanks{M.-J. Kim is with the Department of Artificial Intelligence, Hanyang University, Seoul, South Korea. E-mail:kmj0792@hanyang.ac.kr}
\thanks{Y.-C. Lee is with the School of Computational Science and Engineering, Georgia Institute of Technology, Atlanta, GA, USA. E-mail:yeonchang@gatech.edu}
\thanks{David Y. Kang is with the School of Information, University of Michigan,
Ann Arbor, MI, USA. E-mail:dyskang@umich.edu}
\thanks{S.-W. Kim is with the Department of Computer Science, Hanyang University, Seoul, South Korea. E-mail:wook@hanyang.ac.kr}
\thanks{\small\IEEEauthorrefmark{1}Two first authors have contributed equally to this work.}
\thanks{\small\IEEEauthorrefmark{3}Corresponding author.}
}

\maketitle

\newcommand{\spec}{{\it spec.}}
\newcommand{\aka}{{\it a.k.a.}}
\newcommand{\ie}{{\it i.e.}}
\newcommand{\eg}{{\it e.g.}}
\newcommand{\ours}{TrustSGCN}

\newcommand{\egonets}{\texttt{EgoNets}}
\newcommand{\egonet}{\texttt{EgoNet}}
\newcommand{\tgcn}{\texttt{T\mbox{-}GCN}}
\newcommand{\utgcn}{\texttt{U\mbox{-}GCN}}
\newcommand{\fextra}{\texttt{FExtra}}
\newcommand{\uniform}{\texttt{Uniform}}
\newcommand{\reverse}{\texttt{Reverse}}
\newcommand{\mean}{\texttt{Mean}}
\newcommand{\sampler}{\texttt{Sampler}}
\newcommand{\all}{\texttt{All}}

\newcommand\red[1]{\textcolor{red}{#1}}
\newcommand\green[1]{\textcolor{green}{#1}}
\newcommand\blue[1]{\textcolor{blue}{#1}}
\newcommand\violet[1]{\textcolor{violet}{#1}}
\newcommand\change[1]{\textcolor{magenta}{#1}}
\newcommand\add[1]{\textcolor{blue}{#1}}

\newcommand{\yc}[1]{\textcolor{blue}{[YC: #1]}}
\newcommand{\mj}[1]{\textcolor{blue}{[MJ: #1]}}
\newcommand{\dk}[1]{\textcolor{red}{[DK: #1]}}

\begin{abstract}
The problem of representing nodes in a signed network as low-dimensional vectors, known as signed network embedding (SNE), has garnered considerable attention in recent years.
While several SNE methods based on \emph{graph convolutional networks} (GCN) have been proposed for this problem, 
we point out that they significantly rely on the assumption that the decades-old \emph{balance theory} always holds in the real-world. 
To address this limitation, we propose a novel GCN-based SNE approach, named as \ours, which corrects for incorrect embedding propagation in GCN by utilizing the trustworthiness on edge signs for high-order relationships inferred by the balance theory.
The proposed approach consists of three modules: (M1) generation of each node's extended ego-network; (M2) measurement of trustworthiness on edge signs; and (M3) trustworthiness-aware propagation of embeddings.
Furthermore, \ours\ learns the node embeddings by leveraging two well-known societal theories, \textit{i.e.}, balance and status. 
The experiments on four real-world signed network datasets demonstrate that \ours\ consistently outperforms five state-of-the-art GCN-based SNE methods. 
The code is available at \href{https://github.com/kmj0792/TrustSGCN}{https://github.com/kmj0792/TrustSGCN}.
\end{abstract}


\IEEEpeerreviewmaketitle

\section{Introduction}\label{sec:introduction}
\noindent\textbf{Background.} In real-world networks, interactions between nodes are often \emph{signed}, representing two contrasting relationships, such as friend/enemy and support/oppose~\cite{TangCAL16, YuanWX17, DoreianM15}. 
With the advent of signed networks, we can better understand the complex relationships between nodes and measure the polarization in social discussions~\cite{alsinet2021measuring,askarisichani2019structural,BonchiGGOR19,lee2018goccf, HwangPKLL16,KoRHJKPHTK23}. 
Motivated by such broad applicability, the problem of signed network embedding (SNE)~\cite{WangTACL17}, which aims to represent nodes in a given signed network as low-dimensional vectors, has increasingly attracted attention in recent years \cite{LeeSHK20,XuZLY00H22}. 
The learned vectors (\ie, embeddings) can be used as intrinsic features of nodes in solving various downstream tasks related to information retrieval and data mining, including link prediction \cite{HuangSS22,YooLSK22,WangZHXGL18} and recommendation \cite{LeeLLK21,HuiZZWN22,TangAL16,KimYSLIK23}.

In particular, with the advance of \emph{graph convolutional networks} (GCN) \cite{NiepertAK16,KipfW16,lee2022thor}, many GCN-based SNE methods have been proposed recently \cite{Derr0T18,HuangSHC19,LiTZC20,HuangSHC21,ShuDC0ZXS21}. 
They design GCNs that can model the \emph{homophily} nature of positive relationships and the \emph{heterophily} nature of negative relationships in a given signed network.
Furthermore, they leverage the well-known \emph{balance theory}~\cite{cartwright1956structural,heider1946attitudes} in social sciences to understand the formation of high-order signed relationships in the signed network.
It states that such high-order relationships tend to form \emph{balanced triads} that satisfy the following rules: ``a friend of my friend is my friend,'' ``a friend of my enemy is my enemy,'' ``an enemy of my friend is my enemy,'' and ``an enemy of my enemy is my friend.''

\begin{figure}[t]
\centering
\includegraphics[width=0.93\linewidth]{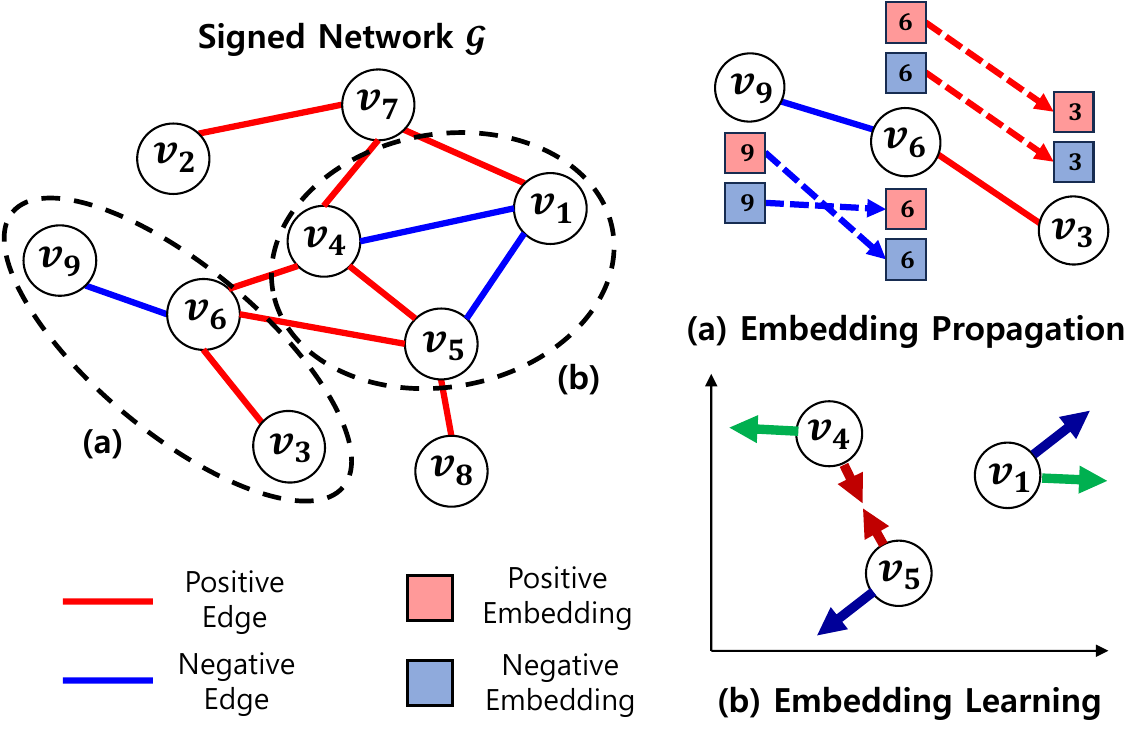}
\vspace{-0.3cm}
\caption{The use of balance theory in existing GCN-based SNE methods.} \label{fig:SNE}
\vspace{-0.4cm}
\end{figure}

To take advantage of the balance theory, most GCN-based SNE methods generate two types (\ie, positive and negative) of embeddings for each node $v_i$ by performing the following embedding propagation~\cite{Derr0T18,HuangSHC19,LiTZC20,HuangSHC21,ShuDC0ZXS21}: embeddings of the \emph{same (resp. opposite) polarity} are propagated between $v_i$ and its neighbors \emph{with positive (resp. negative) relationships} (Figure~\ref{fig:SNE}-(a)). Then, they learn the node embeddings so that the likelihood of existent edges in the input signed network can be maximized.
Along with this optimization, a more recent work~\cite{HuangSHC21,kang2021adversarial} additionally learns the node embeddings such that the proximities between \emph{three nodes} incident to the balanced triads can be preserved (Figure~\ref{fig:SNE}-(b)).
To sum up, existing GCN-based SNE methods significantly rely on the balance theory when they generate and learn node embeddings.

\vspace{1mm}
\noindent\textbf{Motivation.} However, \ul{edge relationships in real settings of signed networks often violate the rules of the balance theory~\mbox{\cite{LeeLLK21,Estrada19}}}. 
For instance, on platforms like Reddit\footnote{https://www.reddit.com}, a social-media discussion platform, it is common to observe situations where individuals have conflicting opinions on different topics, \eg, even though Bob supports Alice who opposes John on a specific topic (\eg, politics), Bob may support John on a different topic (\eg, sports).
Additionally, in the context of politics, it often occurs in the United States House of Representatives that members belonging to a particular political party have no negative opinions towards bills or speeches initiated by members of the opposing party. 

As a demonstration, Table~\ref{table:balance} shows that a high number of \emph{unbalanced triads}, which do not satisfy the balance theory, appear in the real-world signed network datasets.
For instance, on Epinions, we can see that 71\% of triads with the prior signs of (+,-) have a posterior sign of +,
indicating ``an enemy of my friend is my friend.'' It indicates that \ul{the balance theory can bring about substantial errors in predicting the edge signs for high-order relationships when blindly employed.}
Consequently, the node embeddings, which learn the \emph{incorrect edge signs} inferred by the balance theory, can adversely affect the downstream tasks.

\vspace{1mm}
\noindent\textbf{Our Work.} In this work, we aim to measure the trustworthiness on edge signs for high-order relationships inferred by balance theory and correct wrong embedding propagation based on the trustworthiness thus measured. Toward this goal, we propose a novel SNE approach, named as 
\ours, which learns \underline{Trust}worthiness on edge signs for \underline{S}igned \underline{G}raph \underline{C}onvolutional \underline{N}etworks. For each node $v_i$ in a given signed network, \ours\ first constructs $v_i$'s extended ego-network consisting of $v_i$'s direct and high-order neighbors whose trustworthiness on edge signs will be measured. With the ego network, \ours\ validates whether the edge signs for high-order neighbors inferred by balance theory are \emph{trustworthy or not}. Then, \ours\ performs \emph{different embedding propagation} according to trustworthy or untrustworthy edge signs, updating $v_i$'s 
positive and negative embeddings.

\begin{table}[t]
\footnotesize
\caption{Ratios of balanced/unbalanced triads. Given two signs (\textit{i.e.}, prior signs) in a triad, we check whether the remaining sign (\textit{i.e.}, posterior sign) satisfies the balance theory.}
\vspace{-0.2cm}
\label{table:balance}
\resizebox{.49\textwidth}{!}{
\begin{tabular}{cc|cc|cc|cc}
\toprule
\multirow{2}{*}{\textbf{Triads}}  & \textbf{Prior} & \multicolumn{2}{c|}{\textbf{$(+,+)$}} &  \multicolumn{2}{c|}{\textbf{$(+,-)$/$(-,+)$}} & \multicolumn{2}{c}{\textbf{$(-,-)$}} \\ 
& \textbf{Posterior} & $+$ & $-$ & $+$ & $-$ & $+$ & $-$ \\ \midrule
\multirow{4}{*}{\textbf{Datasets}} & \textbf{Bitcoin-Alpha} & 85\% & 15\% & 63\% & 37\% & 93\% & 7\% \\
& \textbf{Bitcoin-OTC} & 86\% & 14\% & 42\% & 58\% & 94\% & 6\% \\
& \textbf{Slahsdot} & 82\% & 18\% & 51\% & 49\% & 79\% & 21\%  \\
& \textbf{Epinions} & 96\% & 4\% & 71\% & 29\% & 90\% & 10\%  \\ \midrule
\multicolumn{2}{c|}{\textbf{Balanced?}} & $\bigcirc$ & \large$\times$ & \large$\times$ & $\bigcirc$ & $\bigcirc$ & \large{$\times$}\\
\bottomrule
\end{tabular}
}
\vspace{-0.2cm}
\end{table}

Finally, \ours\ learns the node embeddings to preserve the proximity between nodes connected by existing edges in the embedding space. To achieve this, we incorporate not only the balance theory but also another social theory, \ie, the \emph{status theory}~\cite{leskovec2010signed}, into the learning mechanism of \ours. We experimentally demonstrate that (1) considering the trustworthiness on edge signs helps accurately preserve the proximities between nodes, (2) \ours\ outperforms five state-of-the-art GCN-based SNE methods~\cite{Derr0T18,HuangSHC19,LiTZC20,HuangSHC21,ShuDC0ZXS21} in terms of an edge sign prediction task, and
(3) \ours\ requires a reasonable training time to achieve higher accuracy than competitors.

\vspace{1mm}
\noindent\textbf{Contributions.} Our contributions are summarized as follows:
\begin{itemize}[leftmargin=*]
\item \textbf{Important Observation}: We point out that the balance theory could cause incorrect embedding propagation under existing GCN-based SNE methods, if relied blindly on.
\item \textbf{Effective SNE Approach}: We propose \ours\ that learns node embeddings based on trustworthiness on edge signs for signed graph convolutional networks.

\begin{itemize}
    \item We design three modules to facilitate embedding propagation, taking into account the trustworthiness on edge signs for high-order signed relationships.
    \item We design a loss function to facilitate embedding learning while preserving the signed and directed relationships based on the balance and status theories.
\end{itemize}
\item \textbf{Comprehensive Validation}: Our comprehensive validation demonstrates the effectiveness of each individual design choice in \ours, as well as the superiority of our approach achieved when all these choices are combined.
\end{itemize}

The earlier version of this manuscript was presented as a short paper (\ie, conference version)~\cite{trustsgcn2023} at ACM SIGIR 2023, featuring initial ideas and preliminary experimental results. 
In this extended version, we (1) provide comprehensive and detailed descriptions of our motivation, intuitions behind our ideas, and the implementation details of the proposed approach, (2) design a multi-objective loss function that jointly considers the balance and status theories, and (3) conduct more experiments to demonstrate the effectiveness of the proposed approach more clearly.

\vspace{1mm}
\noindent\textbf{Organization.} 
The rest of this paper is organized as follows.
In Section~\ref{sec:related_work}, we review previous studies on the SNE problem.
In Section~\ref{sec:approach}, we present our proposed approach in detail.
In Section~\ref{sec:evaluation}, we validate the effectiveness of the proposed approach through extensive experiments. 
Finally, we summarize 
the paper in Section~\ref{sec:conclusions}.

\section{Related Work}\label{sec:related_work}
In this section, we briefly review the social theories commonly used in signed network analysis (Section~\ref{sec:social_thoery}), and then discuss how existing SNE methods take advantage of these social theories (Section~\ref{sec:existing_methods}).

\vspace{-0.2cm}

\subsection{Social Theories for Signed Networks}\label{sec:social_thoery}

\noindent\textbf{Balance Theory}. 
The balance theory~\cite{cartwright1956structural,heider1946attitudes} is a concept derived from social psychology that provides insights into the intricate dynamics of human relationships.
It offers an explanation for the signed relationships between three nodes in a signed undirected network.

\begin{figure}[t]
\centering
\includegraphics[width=0.98\linewidth]{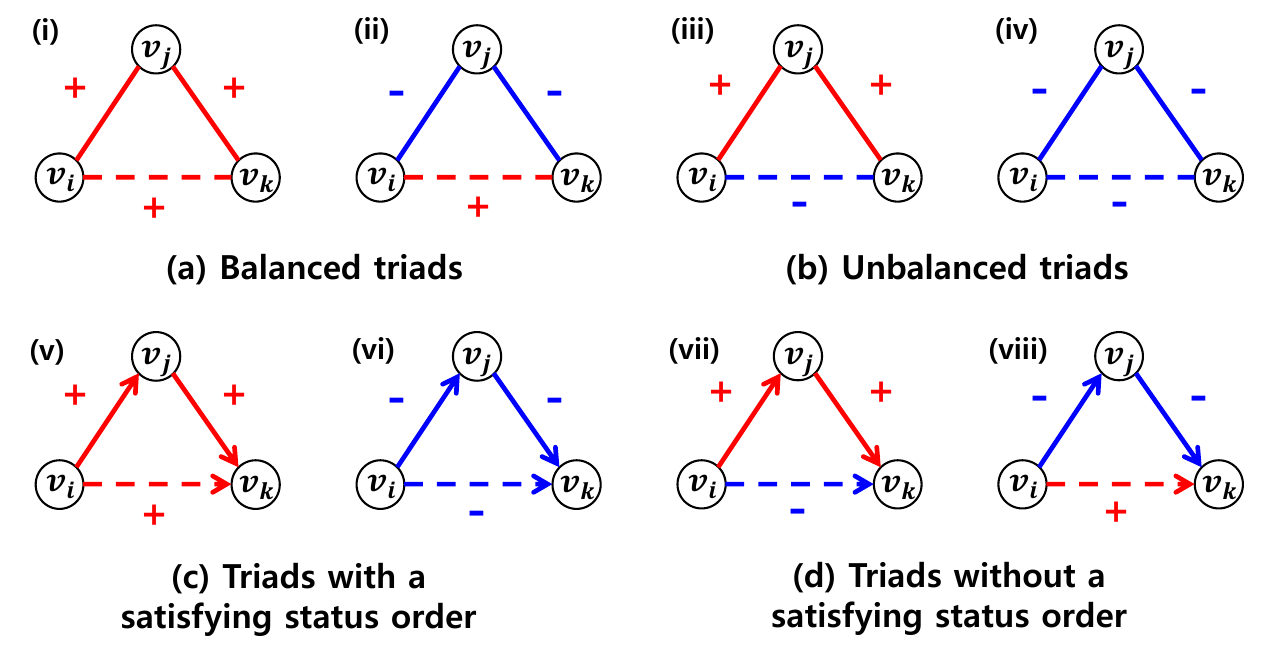}
\vspace{-0.3cm}
\caption{Examples of triads where balance and status theories are followed and not followed}\label{fig:theories}
\end{figure}

According to the balance theory, an unknown signed relationship between two nodes $v_i$ and $v_k$ can be inferred from the two signed relationships known between each of the nodes and their common neighbor $v_j$.
This inference is guided by the following rules~\cite{cartwright1956structural,heider1946attitudes}: 
(1) a friend of my friend is my friend; (2) a friend of my enemy is my enemy; (3) an enemy of my friend is my enemy; and (4) an enemy of my enemy is my friend. 
In a signed network, triads that adhere to these rules are called `\textit{balanced triads}' (\ie, having an even number of negative edges), while those that deviate from these rules are called `\textit{unbalanced triads}' (\ie, having an odd number of negative edges). Figures~\ref{fig:theories}-(a) and (b) illustrate the distinction between balanced and unbalanced triads. 
For instance, in Figure~\ref{fig:theories}-(i), $v_k$, who is a friend of $v_i$'s friend $v_j$, is considered as a friend of $v_i$, making it a balanced triad.
Conversely, in Figure~\ref{fig:theories}-(iii), 
$v_k$ is considered as an enemy of $v_i$, which violates the balance theory. Three nodes $v_i$, $v_j$, and $v_k$ are in an unbalanced triad.
However, it is important to note that in real-world settings of signed networks, balance theory does not always hold, as mentioned in Section~\ref{sec:introduction} (see Table~\ref{table:balance}).

\vspace{1mm}
\noindent\textbf{Status Theory}. 
The status theory~\cite{leskovec2010signed} is another theory from social psychology and is used to model \emph{signed directed networks}.
It explains the social status of nodes in signed directed networks based on the signs and directions of edges, where status can refer to the reputation, ranking, and skill level.

According to the status theory, an unknown signed and directed relationship between two nodes $v_i$ and $v_k$ can be inferred from the two signed and directed relationships known between each of the nodes and their common neighbors $v_j$.
This inference is guided by the following rules~\cite{leskovec2010signed}: 
(1) when a positive link exists from the source node to the destination node, the source node regards the destination node as a friend with a higher status than itself; and (2) when a negative link exists from the source node to the destination node, the source node regards the destination node as an enemy with a lower status than itself.
In a signed directed network, the triads that adhere to these rules are called `\textit{triads with a satisfying status order}', while those that violate these rules are called `\textit{triads without a satisfying status order}.'
Figures~\ref{fig:theories}-(c) and (d) illustrate the distinction between triads with and without a satisfying status order. 
As in Figure~\ref{fig:theories}-(v) and Figure~\ref{fig:theories}-(vii), suppose $v_j$ is a friend of $v_i$ with a higher status than $v_i$, and $v_k$ is a friend of $v_j$ with a higher status than $v_j$.
In this case, $v_k$ is considered as a friend and an enemy of $v_i$ with a higher and lower status than $v_i$ in Figure~\ref{fig:theories}-(v) and Figure~\ref{fig:theories}-(vii), respectively;
thus, the former and the latter represent a triad with and without a satisfying status order, respectively.

\subsection{Signed Network Embedding Methods}\label{sec:existing_methods}
SNE methods aim to represent nodes in a given signed network as low-dimensional vectors so that the vectors preserve structural and semantic properties in the network.

\vspace{1mm}
\noindent\textbf{Traditional SNE Methods.} 
SiNE\cite{wang2017signed}, SIDE\cite{kim2018side}, and BESIDE\cite{chen2018bridge} are traditional methods that leverage balance theory to learn the signed relationships between nodes.
They ensure that the nodes with a positive relationship are positioned closer to each other in the embedding space, while the nodes with a negative relationship are positioned farther apart.
In addition, SIDE employs random walks based on balance theory to exploit the unknown signed relationships between nodes, while BESIDE models and learns ``bridge'' edges whose adjacent nodes have no common neighbors based on status theory.

\vspace{1mm}
\noindent\textbf{GCN-based SNE Methods.} 
Several SNE methods have been developed \cite{wang2017signed, kim2018side, chen2018bridge}, incorporating successful convolution-based techniques \cite{Derr0T18, LiTZC20, liu2022lightsgcn, HuangSHC19, HuangSHC21, ShuDC0ZXS21} in representation learning.
First, SGCN\cite{Derr0T18}, SNEA\cite{LiTZC20}, and LightSGCN\cite{liu2022lightsgcn} utilize balance theory for predicting high-order signed relationships and designing their propagation strategies. 
As mentioned in Section~\ref{sec:introduction}, they generate positive and negative embeddings for each node so that the embeddings of the same (resp. opposite) polarity are propagated between the nodes with positive (resp. negative) relationships.
Note that SGCN is the first GCN-based SNE method. 
Then, SNEA adds a self-attention mechanism to SGCN, while LightSGCN simplifies the embedding propagation process by using a linear approach.

On the other hand, SIGAT\cite{HuangSHC19} and SDGNN\cite{HuangSHC21} jointly consider the sign and direction information of signed networks. 
They define motifs (\eg, $\Delta_{i,j,k}$: $v_i\rightarrow^+ v_j$, $v_i\rightarrow^+ v_k$, $v_k\rightarrow^+ v_j$) based on balance and status theories, and utilize motif-based Graph Attention Networks (GAT) for embedding propagation.
Furthermore, SDGNN introduces loss functions that incorporate sign, direction, and triangle information.
SGCL\cite{ShuDC0ZXS21} enhances the embedding generation process with graph augmentations based on balance theory and additionally models the augmented signed relationships between nodes by using contrastive learning.

In summary, while most existing SNE methods successfully incorporate balance theory into their design, they face the challenge of learning the incorrect edge signs predicted by the balance theory.

\begin{figure*}[t]
\centering
\vspace{-0.2cm}
\includegraphics[width=0.95\linewidth]{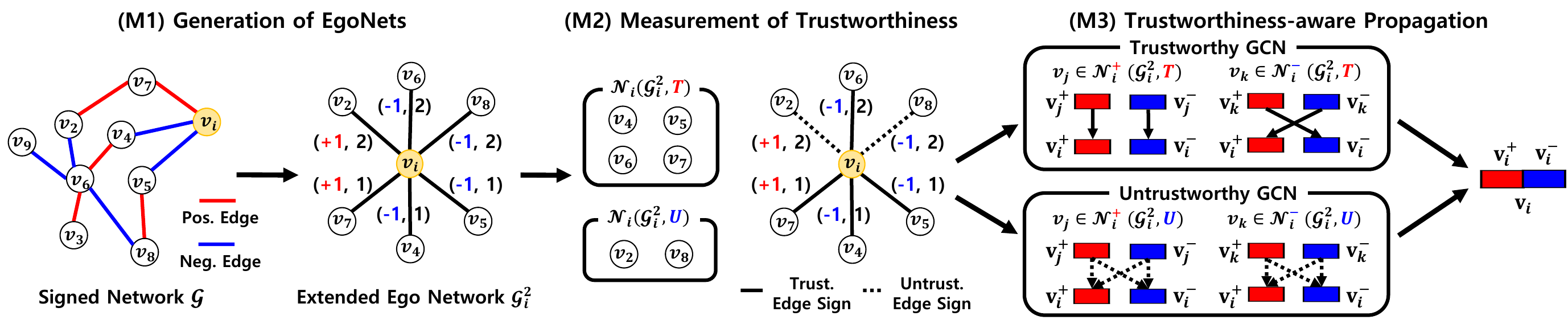}
\vspace{-0.3cm}
\caption{Overview of \ours, which performs embedding propagation by considering trustworthiness on edge signs.} \label{fig:overview}
\vspace{-0.4cm}
\end{figure*}

\section{The Proposed Approach: \ours}\label{sec:approach}
In this section, we describe \ours, a novel SNE approach that learns node embeddings based on the trustworthiness on edge signs.
We first formulate the problem of SNE and present an overview of \ours.
The SNE problem is formulated as follows. 
Let $\mathcal{G}=(\mathcal{V}, \mathcal{E}^+,\mathcal{E}^-)$ be a given signed network, where $\mathcal{V}=\left\{v_1,  v_2, \cdots, v_l\right\}$ denotes the set of $l$ nodes, and $\mathcal{E}^+$ and $\mathcal{E}^-$ represent the sets of positive and negative edges, respectively.
SNE methods aim to learn a function $\mathbf{f}:\mathcal{V}\rightarrow \mathbb{R}^\textit{d}$, which maps each node $v_{i}\in\mathcal{V}$ to a \textit{d}-dimensional embedding.

\ours\ generates the node embeddings via three key modules (see Figure~\ref{fig:overview}): (M1) generation of extended ego networks (\egonets), (M2) measurement of trustworthiness on edge signs, and (M3) trustworthiness-aware propagation of embeddings. 
Then, \ours\ learns them through a loss function based on the balance and status theories.
Table~\ref{table:notation} summarizes a list of notations used in this paper.

\begin{table}[t]
\small
\centering
\caption{Notations used in this paper}
\vspace{-0.3cm}
\label{table:notation}
\resizebox{.49\textwidth}{!}{
 \renewcommand{\arraystretch}{1.5}
\begin{tabular}{c|l}
\toprule
\textbf{Notation} & \multicolumn{1}{c}{\textbf{Description}}\\ \midrule
$\mathcal{G}$ & Signed network  \\ 
$\mathcal{V}$ & Set of nodes   \\ 
$\mathcal{E}, \mathcal{E}^{+}, \mathcal{E}^{-}$   & Sets of edges, positive edges, and negative edges  \\ 
\multirow{2}{*}{$\mathbf{v}_i $, $\mathbf{v}_i^+ $, $\mathbf{v}_i^-$}  & Embedding, positive embedding, negative embedding
 \vspace{-0.1cm}\\ & of $v_i$ 
\\ 
$\mathbf{m}^+$, $\mathbf{m}^-$ & Positive and negative message
\\
 $d$ & Dimensionality of embeddings 
 \\
 
\midrule
\multirow{2}{*}{$\mathcal{G}_i^n$} & $v_i$'s EgoNet with neighbors of $v_i$ whose path lengths \vspace{-0.1cm}\\ &  from $v_i$ are less than or equal to $n$ in $\mathcal{G}$
\\

\multirow{2}{*}{$\mathcal{N}_i^n(\mathcal{G})$ \vspace{-0.1cm}}& Set of $v_i$'s neighbors whose path lengths from $v_i$ are \vspace{-0.1cm}\\ &  less than or equal to $n$ in $\mathcal{G}$
\\  

\multirow{2}{*}{$e_{ij}=(s_{ij}, p_{ij})$} & Label for each edge in $\mathcal{G}_i^n$, including a pair of an \vspace{-0.1cm}\\ & edge sign $s_{ij}$ and a path length $p_{ij}$

\\
\multirow{2}{*}{$\hat{s}_{ij}, Trust(\hat{s}_{ij})$} & Edge sign between $v_i$ and $v_j$ and its trustworthiness \vspace{-0.1cm}\\ & inferred by FExtra 
\\

\multirow{2}{*}{$\mathcal{N}_i(\mathcal{G}_i^n, T)$, $\mathcal{N}_i(\mathcal{G}_i^n, U)$} & Sets of $v_i$'s neighbors with trustworthy and untrust- \vspace{-0.1cm}\\ & worthy edge signs in $\mathcal{G}_i^n$
\\


\multirow{2}{*}{$\mathcal{N}_i^{+}(\mathcal{G}_i^n, *)$, $\mathcal{N}_i^{-}(\mathcal{G}_i^n, *)$} & Sets of $v_i$'s neighbors with $s_{ij}=+1$ and $s_{ij}=-1$ \vspace{-0.1cm}\\ &in $\mathcal{G}_i^n$
\\


 \midrule
$H$ & Number of \tgcn\ and \utgcn\ layers
\\
\multirow{2}{*}{$r_{\left({ a},{ b},{c} \right)}$}  & Ratio of propagation according to two prior signs
\vspace{-0.1cm}\\ &  (\ie, {a} and {b}) and a posterior sign (\ie, {c})
\\ 
$n$ & Maximum path length between two nodes to be included in the EgoNet
\\
$\alpha_{p_{ij}}$ &  Attention weight according to $p_{ij}$
\\
$\beta$ & Pre-defined threshold for $Trust(\hat{s}_{ij})$
\\
$\gamma$ & Number of randomly sampled nodes 
\\
 $\lambda$ & Weight for the status loss
 \\
\bottomrule
\end{tabular}
}
\vspace{-0.45cm}
\end{table}

\subsection{Key Modules}\label{sec:modules}
For each node $v_i$ in a given signed network $\mathcal{G}$, \ours\ randomly generates $v_i$'s (initial) positive and negative embeddings $\mathbf{v}_i^+ $, $\mathbf{v}_i^-$ and performs the following three modules sequentially.

\vspace{1mm}
\noindent{\bf(M1) Generation of EgoNets.} 
We sample a set of nodes $v_j$ among $v_i$'s indirect neighbors in $\mathcal{G}$,
where the trustworthiness on the edge signs between $v_i$ and $v_j$ predicted by balance theory will be measured (\ie, M2) and the trustworthiness-based propagation will be conducted (\ie, M3).
To this end, we generate $v_i$'s \egonet\ consisting of edges between $v_i$ and $v_j$ (\ie, $v_i$'s direct and $n$-hop neighbors in $\mathcal{G}$).
The intuition behind this design choice is that since the prediction accuracy on edge signs by balance theory decreases rapidly as the length of the path between two nodes increases~\cite{LeeLLK21}, we only use nodes $v_j$, whose paths from $v_i$  
are short, as $v_i$'s neighbors in \egonet. 
That is, a $v_i$'s \egonet\ $\mathcal{G}_i^n$ can be defined:
\begin{equation}
    \mathcal{G}_i^n=\left\{(v_i, e_{ij}, v_j)|v_j\in \mathcal{N}_i^n(\mathcal{G}), e_{ij}=(s_{ij}, p_{ij})\right\},
\label{eq:egonet}
\end{equation}
where $\mathcal{N}_i^n(\mathcal{G})$ represents a set of nodes $v_j$ whose path length from $v_i$ is less than or equal to \textit{n} in $\mathcal{G}$; also, $e_{ij}$ indicates an edge label between $v_i$ and $v_j$, including a pair of an edge sign $s_{ij} \in \left\{+1, -1\right\}$ and a path length $p_{ij}$. 
Here, for $s_{ij}$ between $v_i$ and its indirect neighbors $v_j$ in $\mathcal{G}$, we assign the edge signs predicted by the balance theory. 

For instance, consider $v_2$ and $v_6$ in Figure~\ref{fig:overview}, which are 2-hop neighbors of $v_i$.
In this case, we can consider $v_2$ (resp. $v_6$) as the friend (resp. enemy) of $v_i$ based on the balance theory, since there exists an even (resp. odd) number of negative edges between $v_2$ (resp. $v_6$) and $v_i$. 
Consequently, the edge signs $s_{i2}$ and $s_{i6}$ in $\mathcal{G}_i^n$ are assigned positive and negative, respectively.
Moreover, when there are multiple paths from $v_i$ to $v_j$ in $\mathcal{G}$ with lengths less than or equal to \textit{n}, the \egonet\ $\mathcal{G}_i^n$ of $v_i$ takes the form of a \emph{multigraph}~\cite{BodenGHS12,IngalalliIP18}, which allows multiple edges between the same pair of nodes. 
However, note that if there is a direct edge between $v_i$ and $v_j$, longer paths are discarded. 
The intuition behind this is to avoid the noise arising from uncertain high-order relationships when we have a \textit{reliable} direct edge.

Suppose that we construct $v_i$'s \egonet\ $\mathcal{G}_i^3$ by using its 3-hop neighbors (\ie, $n=3$). 
In this scenario, there are three distinct paths between $v_i$ and $v_6$ in Figure~\ref{fig:overview}, \ie, ($v_i$,$v_4$,$v_6$), ($v_i$,$v_7$,$v_2$,$v_6$), and ($v_i$,$v_5$,$v_8$,$v_6$).
As a result, $\mathcal{G}_i^3$ includes three edges between $v_i$ and $v_6$.
Since these edges may have different labels, one can raise concerns about the cases where the same pair of nodes exhibit both positive and negative signs through different paths.
However, \ours\ can address such a contradictory case by not only measuring the trustworthiness of each edge sign in (M2) but also prioritizing more frequent propagation of the majority sign in (M3).

\vspace{1mm}
\noindent{\bf(M2) Measurement of Trustworthiness on Edge Signs.} 
We measure the trustworthiness on the edge signs $s_{ij}$ in $\mathcal{G}_i^n$ generated by (M1). 
Note that $\mathcal{G}_i^n$ includes both \emph{existent} and \emph{non-existent} edges in the input signed network $\mathcal{G}$:
the existent edges have \emph{actual signs} expressed in $\mathcal{G}$, while the non-existent edges have \emph{inferred signs} by balance theory. 
In this sense, we assume that the actual signs are all trustworthy.
On the other hand, since the balance theory usually provides inaccurate predictions on edge signs
(see Table~\ref{table:balance}), we measure the trustworthiness on the inferred signs.

Accordingly, for the edges with $p_{ij}\geq2$ in $\mathcal{G}_i^n$, we predict their signs in a different way by utilizing the additional \emph{topological information} related to both $v_i$ and $v_j$ (not the combination of the edge signs included in the path from $v_i$ to $v_j$).  
Specifically, we first learn a logistic regression classifier model, FExtra~\cite{LeskovecHK10,LiFZ17}, based on various topological features.
To this end, we first extract 23 features for each node pair ($v_i$, $v_j$) in $\mathcal{G}$, thus constructing a feature vector $\mathbf{f}_{i,j}=\{f_1(i,j),\cdots,f_{23}(i,j)\}$ for the corresponding node pair, where $f_k(i,j)$ indicates $k$-th feature value for the node pair. 
Among the 23 features, 7 features are related to the degrees of $v_i$ and $v_j$.
On the other hand, the remaining 16 features are related to the triads consisting of $v_i$, $v_j$, and their common neighbors in $\mathcal{G}$.
The details for features can be founded in Appendix~\ref{app:feature}.

Using the learned classifier, we then predict the sign, $\hat{s}_{ij} \in \left\{+1, -1\right\}$, of each edge with $p_{ij}\geq2$ in $\mathcal{G}_i^n$ (\ie, non-existent edge in $\mathcal{G}$) based on the following probabilities:

\vspace{-0.1cm}
\footnotesize
\begin{equation}
    \begin{aligned}
    P(+|\mathbf{f}_{i,j}) &= \frac{1}{1+e^{-(b_0+\sum_{k=1}^{23} b_kf_k(i,j))}}, \\
    P(-|\mathbf{f}_{i,j}) &= 1-P(+|\mathbf{f}_{i,j}),
    \end{aligned}
\label{eq:FExtra}
\end{equation}
\normalsize
where $P(+|\mathbf{f}_{i,j})$ and $P(-|\mathbf{f}_{i,j})$ represent the probability that the edge sign between $v_i$ and $v_j$ is positive and negative, respectively. 
Also, $b_0$ indicates a bias coefficient, and $b_1,\cdots,b_{23}$ indicate coefficients for each feature, all of which are estimated based on training data~\cite{LeskovecHK10,LiFZ17}.
Meanwhile, the classifier also outputs the trustworthiness, $Trust(\hat{s}_{ij})$, on the inferred edge sign $\hat{s}_{ij}$. 
Note that we perform this sign prediction task as a pre-processing task in advance to avoid the computational overhead for the training of \ours.

Based on $\hat{s}_{ij}$ and $Trust(\hat{s}_{ij})$, we now decide whether each edge sign $s_{ij}$ predicted by balance theory is trustworthy or not.  
To this end, we employ the following two conditions~\cite{LeeLLK21}:
\begin{itemize}[leftmargin=*]
    \item (C1) $Trust(\hat{s}_{ij}) > \beta$, where $\beta$ represents a pre-defined threshold for $Trust(\hat{s}_{ij})$
    \item (C2) $\hat{s}_{ij} = {s}_{ij}$
\end{itemize}
That is, we consider $s_{ij}$ predicted by balance theory as trustworthy if both conditions are satisfied, otherwise untrustworthy.
In other words, when $Trust(\hat{s}_{ij})$ gets higher than a pre-defined threshold and the edge signs predicted by FExtra and by balance theory are identical, we finally trust $s_{ij}$ predicted by balance theory.
By doing so, we can classify $v_i$'s neighbors $v_j$ in $\mathcal{G}_i^n$ into two subsets: a set $\mathcal{N}_i(\mathcal{G}_i^n,{T})$ of $v_j$ with \emph{trustworthy} edge signs and another set $\mathcal{N}_i(\mathcal{G}_i^n,{U})$ of $v_j$ with \emph{untrustworthy} edge signs.
Also, each subset can be divided according to the edge signs $s_{ij}$, \ie, $\mathcal{N}_i^{+}(\mathcal{G}_i^n,*)$ and $\mathcal{N}_i^{-}(\mathcal{G}_i^n,*)$.

\vspace{1mm}
\noindent{\bf(M3) Trustworthiness-aware Propagation of Embeddings.} 
We perform different embedding propagation according to the subsets obtained from (M2). 
To this end, we devise two types of GCN: (1) trustworthy GCN (in short, \tgcn) propagates the embeddings of nodes in $\mathcal{N}_i (\mathcal{G}_i^n,T)$ to $v_i$; (2) untrustworthy GCN (in short, \utgcn) propagates the embeddings of nodes in $\mathcal{N}_i (\mathcal{G}_i^n,U)$ to $v_i$.

Specifically, \tgcn\ generates positive and negative messages, $\mathbf{m}_{i\leftarrow T}^+(h)$ and $\mathbf{m}_{i\leftarrow T}^-(h)$, of nodes $v_j \in \mathcal{N}_i^+(\mathcal{G}_i^n,T)$ and $v_k \in \mathcal{N}_i^-(\mathcal{G}_i^n,T)$ with trustworthy edge signs, which will be propagated to $v_i$:

\vspace{-0.1cm}
\scriptsize

\begin{equation}
    \begin{aligned}
    \mathbf{m}_{i\leftarrow T}^+(h) = \sigma \bigg( \mathbf{W}^+(h) \bigg(\sum_{v_j \in \hat{\mathcal{N}}_i^+ (\mathcal{G}_i^n,T)} &\alpha_{p_{ij}} \mathbf{v}_j^+(h-1) \\
    &+ \sum_{v_k \in \hat{\mathcal{N}}_i^- (\mathcal{G}_i^n,T)} \alpha_{p_{ik}} \mathbf{v}_k^-(h-1) \bigg)\bigg),\\
    \mathbf{m}_{i\leftarrow T}^-(h) = \sigma \bigg( \mathbf{W}^-(h) \bigg(\sum_{v_j \in \hat{\mathcal{N}}_i^+ (\mathcal{G}_i^n,T)} &\alpha_{p_{ij}} \mathbf{v}_j^-(h-1) \\
    &+ \sum_{v_k \in \hat{\mathcal{N}}_i^- (\mathcal{G}_i^n,T)} \alpha_{p_{ik}} \mathbf{v}_k^+(h-1) \bigg)\bigg),
    \end{aligned}
\label{eq:tgcn}
\end{equation}
\normalsize
where $\hat{\mathcal{N}}_i^+ (\mathcal{G}_i^n,T)$ and $\hat{\mathcal{N}}_i^- (\mathcal{G}_i^n,T)$ represent sets of randomly sampled $\gamma$ nodes from $\mathcal{N}_i^+ (\mathcal{G}_i^n,T)$ and $\mathcal{N}_i^- (\mathcal{G}_i^n,T)$, respectively
Note that random sampling is performed for each node $v_i$ to achieve efficient learning~\cite{chen2018fastgcn} and mitigate the issues of over-smoothing~\cite{hamilton2017inductive} and over-fitting~\cite{kipf2016semi}, which will be validated in Section~\ref{sec:results}.
Additionally, $\mathbf{v}_j^+(h-1)$ and $\mathbf{v}_j^-(h-1)$ denote the positive and negative embeddings of node $v_j$ from the $(h-1)$-th layer, respectively. 
Also, $\sigma$ indicates a sigmoid function, and $\mathbf{W}^+(h)$ and $\mathbf{W}^-(h)$ represent the learnable weight matrices of the $h$-th layer. 
Furthermore, $\alpha_{p_{ij}}$ represents the learnable attention, controlling the weight of propagation for $v_j$'s embeddings
according to the path length $p_{ij}$ from $v_i$ to $v_j$. 
Intuitively, 
\tgcn\ propagates the embeddings of the \emph{same} polarity from $v_j \in \hat{\mathcal{N}}_i^+ (\mathcal{G}_i^n,T)$ to $v_i$, while propagating the embeddings of the \emph{opposite} polarity from $v_k \in\hat{\mathcal{N}}_i^- (\mathcal{G}_i^n,T)$ to $v_i$, in the same way as in the existing GCN-based SNE methods \cite{Derr0T18,LiTZC20}.

Next, \utgcn\ generates positive and negative messages, $\mathbf{m}_{i\leftarrow U}^+(h)$ and $\mathbf{m}_{i\leftarrow U}^-(h)$, of nodes $v_j\in\mathcal{N}_i^+(\mathcal{G}_i^n,U)$ and $v_k\in=\mathcal{N}_i^-(\mathcal{G}_i^n,U)$ with untrustworthy edge signs, which will be propagated to $v_i$:

\vspace{-0.1cm}
\scriptsize
\begin{multline*}
    \mathbf{m}_{i\leftarrow U}^+(h) = \\ 
    \sigma \bigg( \mathbf{W}^+(h) \bigg( \sum_{v_j \in \hat{\mathcal{N}}_i^+ (\mathcal{G}_i^n,U)} \!\!\alpha_{p_{ij}} \left( r_{(+++)} \mathbf{v}_j^+(h-1) + r_{(-++)} \mathbf{v}_j^-(h-1) \right) \\
     +\!\sum_{v_k \in \hat{\mathcal{N}}_i^- (\mathcal{G}_i^n,U)} \!\!\alpha_{p_{ik}}\!\!\left( r_{(+-+)} \mathbf{v}_k^+(h-1) + r_{(--+)} \mathbf{v}_k^-(h-1) \right)\!\!\bigg)\!\!\bigg),
\end{multline*}  
     
\begin{multline}
     \mathbf{m}_{i\leftarrow U}^-(h) = \\
     \sigma \bigg( \mathbf{W}^-(h) \bigg( \sum_{v_j \in \hat{\mathcal{N}}_i^+ (\mathcal{G}_i^n,U)} \!\!\alpha_{p_{ij}}\!\!\left( r_{(++-)} \mathbf{v}_j^+(h-1) + r_{(-+-)} \mathbf{v}_j^-(h-1) \right)  \\
    +\!\sum_{v_k \in \hat{\mathcal{N}}_i^- (\mathcal{G}_i^n,U)} \!\!\alpha_{p_{ik}} \left( r_{(+--)} \mathbf{v}_k^+(h-1) + r_{(---)} \mathbf{v}_k^-(h-1) \right)\!\!\bigg)\!\!\bigg),
    \label{eq:utgcn1}
\end{multline}
\normalsize
where $r_{\left({a},{b},{c} \right)}$ represents the ratio at which $v_j$'s or $v_k$'s signed embeddings are partially propagated to $v_i$ 
according to the following three signs: (1) $\mathbf{v}_j^{a}$ or $\mathbf{v}_k^{a}$ 
 (\ie, a sign ${a}$ of $v_j$'s or $v_k$'s embedding to be propagated); (2) $\mathcal{N}_i^{b} (\mathcal{G}_i^n,U) $ (\ie, a sign ${b}$ of the subset that includes $v_j$ or $v_k$); 
and (3) $\mathbf{m}_{i\leftarrow U}^{c}(h)$ (\ie, a sign ${c}$ of the message propagated to $v_i$). 
For determining ratios of propagation, we consider the structural properties inherent to the input network $\mathcal{G}$. Specifically, given two prior signs $a$ and $b$, we leverage the ratio of triads with a posterior sign $c$ in $\mathcal{G}$ (see Table~\ref{table:balance}). 
For example, when the input network $\mathcal{G}$ is Epinions, $r_{(+-+)}$ and $r_{(+--)}$ become 0.71 and 0.29, respectively.
The intuition behind this design choice is that since the edge sign between the two nodes $v_i$ and $v_j\in\hat{\mathcal{N}}^+_i (\mathcal{G}_i^n,U)$ (resp. $v_k\in\hat{\mathcal{N}}^-_i (\mathcal{G}_i^n,U)$) is uncertain, both positive and negative embeddings of $v_j$ (resp. $v_k$) are partially included in both positive and negative messages~\cite{LeeLLK21}. 
By doing so, we can correct the incorrect propagation by referring to the statistics of a given network without the need for hyperparameters.

\begin{figure}[t]
\centering
\vspace{0.0cm}
\includegraphics[width=0.95\linewidth]{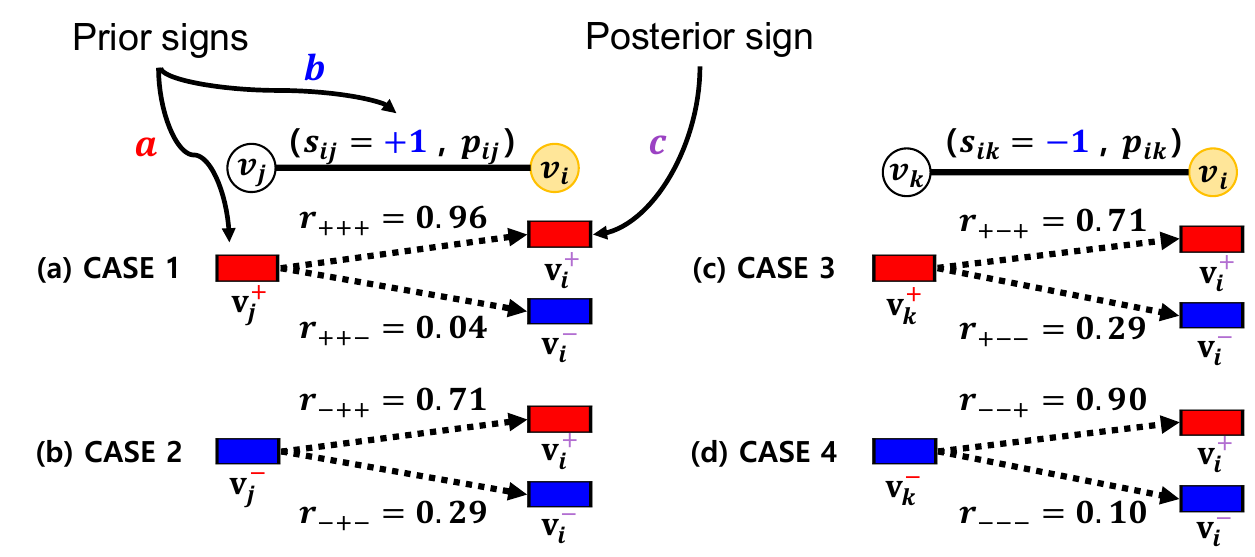}
\vspace{-0.3cm}
\caption{Embedding propagation of untrustworthy GCN on Epinions.}
\label{fig:Propagation}
\vspace{-0.4cm}
\end{figure}

Figure~\ref{fig:Propagation} illustrates the cases where the positive and negative messages of $v_i$ are generated by \utgcn\ on the Epinions dataset.
The cases are divided into four groups according to two prior signs $a$, $b$ 
for two nodes $v_i$ and $v_j/v_k$.
When the inferred edge sign $s_{ij}$ is positive (\ie, $b$ = +), \utgcn\ propagates 96\% (\ie, $r_{(+++)}$) and 4\% (\ie, $r_{(++-)}$) of $v_j$'s positive embedding $\mathbf{v}_j^{+}$ 
to $v_i$'s positive and negative embeddings $\mathbf{v}_i^{+}$ and $\mathbf{v}_i^{-}$, respectively (\ie, CASE 1).
Also, it propagates 71\% (\ie, $r_{(-++)}$) and 29\% (\ie, $r_{(-+-)}$) of $v_j$'s negative embedding $\mathbf{v}_j^{-}$ 
to $\mathbf{v}_i^{+}$ and $\mathbf{v}_i^{-}$, respectively (\ie, CASE 2).
On the other hand, when the inferred edge sign $s_{ik}$ is negative (\ie, $b$ = $-$), \utgcn\ propagates 71\% (\ie, $r_{(+-+)}$) and 29\% (\ie, $r_{(+--)}$) of $v_k$'s positive embedding $\mathbf{v}_k^{+}$ 
to $\mathbf{v}_i^{+}$ and $\mathbf{v}_i^{-}$, respectively (\ie, CASE 3).
Also, it propagates 90\% (\ie, $r_{(--+)}$) and 10\% (\ie, $r_{(---)}$) of $v_k$'s negative embedding $\mathbf{v}_k^{-}$ to $\mathbf{v}_i^{+}$ and $\mathbf{v}_i^{-}$, respectively (\ie, CASE 4). 
It should be noted that \ours\ pre-computes the statistics regarding triads for a given signed network as part of a pre-processing task.

Now, \ours\ updates positive and negative embeddings, $\mathbf{v}_i^+(h)$ and $\mathbf{v}_i^-(h)$, of $v_i$ using the above signed messages as follows:

\vspace{-0.1cm}
\footnotesize
\begin{equation}
    \begin{aligned}
    \mathbf{v}_i^+(h) &= \mathbf{v}_i^+(h-1) + \frac{1}{d_i^+}\left(\mathbf{m}_{i\leftarrow T}^+(h) + \mathbf{m}_{i\leftarrow U}^+(h) \right),\\
    \mathbf{v}_i^-(h) &= \mathbf{v}_i^-(h-1) + \frac{1}{d_i^-}\left(\mathbf{m}_{i\leftarrow T}^-(h) + \mathbf{m}_{i\leftarrow U}^-(h) \right),
    \end{aligned}
\label{eq:aggregation}
\end{equation}
\normalsize
where $d_i^+$ and $d_i^-$ 
represent the numbers of nodes in $\hat{\mathcal{N}}_i^+(\mathcal{G}_i^n,*)$ and $\hat{\mathcal{N}}_i^-(\mathcal{G}_i^n,*)$, respectively.
\ours\ considers $v_i$’s signed embeddings $\mathbf{v}_i^+(H)$ and $\mathbf{v}_i^-(H)$ obtained from the last $H$-th layer as $v_i$’s final signed embeddings $\mathbf{v}_i^+$ and $\mathbf{v}_i^-$. 
Finally, it fuses both signed embeddings into a single embedding $\mathbf{v}_i$, \ie, $\mathbf{v}_i=\mathbf{v}_i^+ ||\mathbf{v}_i^-$.

\subsection{Training}\label{sec:Loss Function}
Given a directed edge ($v_i$, $v_j$) from $v_i$ to $v_j$ in the given signed network $\mathcal{G}$, \ours\ learns the embeddings of both nodes by employing two types of loss functions $\mathcal{L}_{sign}$ and $\mathcal{L}_{status}$.
These loss functions are designed based on social theories, working together to preserve the inherent properties (\ie, sign and direction) of the edge in the embedding space.

First, $\mathcal{L}_{sign}$ aims to learn edge signs based on the balance theory and is formulated as follows~\cite{HuangSHC21,ShuDC0ZXS21,HuangSHC19}:

\vspace{-0.1cm}
\footnotesize
\begin{equation}
    \mathcal{L}_{sign} =\sum_{(v_i, v_j) \in \mathcal{E}} -y_{ij}\log(\sigma(\mathbf{v}_i\mathbf{v}_j)) - (1-y_{ij})\log(1-\sigma(\mathbf{v}_i\mathbf{v}_j)),
\label{eq:signloss}
\end{equation}
\normalsize

where $y_{ij}$ is 1 if the edge sign
is positive, and 0 otherwise.
Intuitively, the goal of this loss is to encourage high proximity between two nodes incident to a positive edge and low proximity between two nodes incident to a negative edge in the embedding space. 
By optimizing this objective, \ours\ naturally improves the preservation of the ternary relationships among the three nodes within balanced triads \cite{WangTACL17, LiTZC20}.

Second, $\mathcal{L}_{status}$ aims to learn edge directions based on the status theory and is formulated as follows \cite{HuangSHC21}:

\vspace{-0.1cm}
\footnotesize
\begin{equation}
    \begin{aligned}
    \mathcal{L}_{status} = \sum_{\left(v_i, v_j\right) \in \mathcal{E}} &-y_{ij}\log{\Big(\sigma\left(s\left(\mathbf{v}_j\right)-s\left(\mathbf{v}_i\right)\right) \Big)} \\ 
    &- \left(1-y_{ij}\right)\log{\Big(\sigma\left(s\left(\mathbf{v}_i\right)-s\left(\mathbf{v}_j\right)\right)\Big)},
    \end{aligned} \label{eq:statusloss}
\end{equation}
\normalsize
where $s\left(\mathbf{v}_i\right)=\sigma\left(\mathbf{W}\cdot \mathbf{v}_i+\mathbf{b}\right)$ indicates the status score of  $v_i$, and $\mathbf{W}$ and $\mathbf{b}$ represent a learnable weight matrix and a bias term, respectively.
Intuitively, guided by the status theory, the goal of this loss is to ensure that, for a positive edge, the destination node's status is higher than that of the source node, while for a negative edge, the destination node's status is lower than that of the source node.

By jointly optimizing the two losses above, the final loss function of \ours\ is defined as follows: 
\vspace{-0.1cm}
\begin{equation}
    \begin{aligned}
    \mathcal{L} &= \mathcal{L}_{sign} + \lambda \mathcal{L}_{status},
    \end{aligned}
\label{eq:7}
\end{equation}
where $\lambda$ indicates a weight parameter for the status loss.
We will analyze the sensitivity of \ours\ to $\lambda$ in Section~\ref{sec:results}.

\section{Evaluation}\label{sec:evaluation}
We designed our experiments, aiming at answering the following key evaluation questions (EQs):
\begin{itemize}[leftmargin=*]
    \item {\textbf{(EQ1})} Does trustworthiness-aware propagation yield effective results for a sign prediction task?
    \item {\textbf{(EQ2})} Does joint learning of two types of loss functions yield effective results for a sign prediction task?
    \item {\textbf{(EQ3})} Does \ours\ achieve superior performance compared to its competitors for a sign prediction task?
    \item {\textbf{(EQ4})} How do different values of a parameter $\beta$ in \ours\ influence the sign prediction accuracy?
    \item {\textbf{(EQ5})} How is the computational overhead (\ie, training time) of \ours\ compared with its competitors?
\end{itemize}
\vspace{-0.2cm}

\subsection{Experimental Setup}\label{sec:setting}

\begin{table}[t]
\footnotesize
\centering
\caption{Dataset statistics}
\vspace{-0.3cm}
\label{table:datasets}
\resizebox{.49\textwidth}{!}{
 \renewcommand{\arraystretch}{1.2}
\begin{tabular}{crrrr}
\toprule
\textbf{Datasets} & \textbf{Nodes} & \textbf{Edges} & \textbf{Positive Edges} & \textbf{Negative Edges} \\ \midrule
\textbf{Bitcoin-Alpha} & 3,784  & 14,145  & 12,729 (89.9\%)  & 1,416 (10.1\%)  \\ 
\textbf{Bitcoin-OTC} & 5,901  & 21,522  & 18,390 (85.4\%)  & 3,132 (14.6\%)  \\ 
\textbf{Slashdot}   & 13,182 & 36,338  & 30,914 (85.1\%)  & 5,424 (14.9\%)  \\ 
\textbf{Epinions}  & 25,148 & 105,061 & 74,060 (70.5\%)  & 31,001 (29.5\%)  \\ 
\bottomrule
\end{tabular}
}
\vspace{-0.45cm}
\end{table}

\noindent\textbf{Datasets}. We used four real-world signed network datasets, which are widely used in previous studies of SNE~\cite{Derr0T18,HuangSHC19,LiTZC20,HuangSHC21,ShuDC0ZXS21}: Bitcoin-Alpha, Bitcoin-OTC, Slashdot, and Epinions. 
The datasets are all publicly available.\footnote{https://snap.stanford.edu/data,  https://www.aminer.cn/data-sna} 
Table~\ref{table:datasets} provides some statistics for the four datasets.

\vspace{1mm}
\noindent\textbf{Competitors}. 
We compared \ours\ with five state-of-the-art GCN-based SNE methods: SGCN \cite{Derr0T18}, SiGAT \cite{HuangSHC19}, SNEA \cite{LiTZC20}, SDGNN \cite{HuangSHC21}, and SGCL \cite{ShuDC0ZXS21}.
For a fair comparison, we set the dimensionality of embeddings to 64 in all methods, following~\cite{Derr0T18, LiTZC20}. 

\vspace{1mm}
\noindent\textbf{Evaluation Task}. 
For testing, we split a dataset into training (80\%) and test (20\%) sets.
Following~\cite{Derr0T18,HuangSHC19,LiTZC20,HuangSHC21,ShuDC0ZXS21}, we employ an edge sign prediction task, which aims to evaluate how accurately each SNE method classifies the edge signs.
To measure the accuracy, we use the three popular metrics: F1-Micro, F1-Macro, and \textit{area under curve} (AUC)~\cite{hanley1982}.

\begin{table*}[!t]
\centering
\vspace{0.0cm}
\caption{Sign prediction accuracies of \ours, \ours(\texttt{T-GCN}), and \ours(\texttt{FExtra})}
\vspace{-0.25cm}
\label{table:eq2-1}
\resizebox{\textwidth}{!} {
 \renewcommand{\arraystretch}{1.5}
\begin{tabular}{c|ccc|ccc|ccc|ccc}
\toprule
\textbf{Datasets} & \multicolumn{3}{c|}{\textbf{Bitcoin-Alpha}} & \multicolumn{3}{c|}{\textbf{Bitcoin-OTC}}  & \multicolumn{3}{c|}{\textbf{Slashdot}} & \multicolumn{3}{c}{\textbf{Epinions}} \\ 
\textbf{Metrics} & \textbf{Micro-F1} & \textbf{Macro-F1} & \textbf{AUC} & \textbf{Micro-F1} & \textbf{Macro-F1} & \textbf{AUC} & \textbf{Micro-F1} & \textbf{Macro-F1} & \textbf{AUC} & \textbf{Micro-F1} & \textbf{Macro-F1} & \textbf{AUC} \\ \midrule

\textbf{\ours} &  \bf{0.921} & \bf{0.721} & \bf{0.867} & \bf{0.901} & \bf{0.773} & {\bf 0.886} & \bf{0.891} & \bf{0.765} & \bf{0.907} & \bf{0.920} & \bf{0.902} & \bf{0.966}  \\ 
\textbf{\ours(\tgcn)} & 0.914 & 0.708 & 0.852 & 0.895 & 0.761 & 0.880 & 0.885 & 0.735 & 0.904 & 0.916 & 0.897 & 0.965  \\ 
\textbf{\ours(\fextra)} & 0.913 & 0.697 & 0.850 & 0.891 & 0.745 & 0.880 & 0.889 & 0.754 & 0.903 & 0.912 & 0.892 & 0.961  \\

\bottomrule
\end{tabular}
}
\vspace{-0.1cm}
\end{table*}

\vspace{1mm}
\noindent\textbf{Implementation Details}. 
All the experiments were conducted on NVIDIA TITAN Xp GPUs with 12GB memory.
We carefully tuned the hyperparameters of competitors and \ours. 
For hyperparameters of competitors, we used the best settings found via grid search. 
For \ours, 
we consistently set $n=3$ (\ie, the number of hops) and $H=1$ (\ie, the number of GCN layers) for all datasets.
In addition, for Bitcoin-Alpha, Bitcoin-OTC, Slashdot, and Epinions, we set $\gamma$ (\ie, the number of randomly sampled nodes) to 30, 30, 20, and 10, $\beta$ (\ie, a pre-defined threshold for $Trust(\hat{s}_{ij})$) to 0.80, 0.95, 1, and 1, and $\lambda$ (\ie, a weight parameter for the status loss) to 1, 0.80, 1, and 1, respectively.

\subsection{Results and Analysis}\label{sec:results}
In this section, we omit some experimental results when we confirmed that the performance of \ours\ for other metrics or parameter settings is consistent with that of the reported results.
The details for the omitted results are available in Appendix~\ref{app:results}.

\begin{figure*}[t!]
\scriptsize
\centering
\begin{tikzpicture}
\centering
\begin{customlegend}[legend columns=2,legend style={align=left,draw=none,column sep=1ex}, legend cell align={left},
  legend entries={\ours, \ours(All)}]
  \addlegendimage{area legend, draw=blue!60, fill=blue!20}
  \addlegendimage{area legend, draw=blue!100, fill=blue!60}
  \end{customlegend}
\end{tikzpicture}

\begin{tikzpicture}
\footnotesize
  \centering
  \begin{axis}[
        ybar=0pt, 
        height=2.9cm, width=5.7cm,
        bar width=7.5pt,
        ymajorgrids=true, 
        tick align=inside,
        major grid style={line width=0.1pt,draw=white},
        enlarge x limits=0.2,
        ymin=0.7, ymax=1,
        axis x line*=left,
        axis y line*=left,
        ylabel={AUC},
        symbolic x coords={
          Bitcoin-Alpha,
          Bitcoin-OTC,
          Slashdot,
          Epinions},
        xtick=data,
        xticklabel style={at={(0.5,-0.5)}, text width=1.5cm, align=center, font=\scriptsize, rotate=30}, %
    ]
    \addplot [draw=blue!60, fill=blue!20] coordinates {
      (Bitcoin-Alpha, 0.867)
      (Bitcoin-OTC, 0.886)
      (Slashdot, 0.907) 
      (Epinions, 0.966)};
    \addplot [draw=blue!100, fill=blue!60] coordinates {
      (Bitcoin-Alpha, 0.841)
      (Bitcoin-OTC, 0.877)
      (Slashdot, 0.878) 
      (Epinions, 0.871)};
  \end{axis}
\end{tikzpicture}
\begin{tikzpicture}
\footnotesize
  \centering
  \begin{axis}[
        ybar=0pt, 
        height=2.9cm, width=5.7cm,
        bar width=7.5pt,
        ymajorgrids=true, 
        tick align=inside,
        major grid style={line width=0.1pt,draw=white},
        enlarge x limits=0.2,
        ymin=0.7, ymax=1,
        axis x line*=left,
        axis y line*=left,
        ylabel={Micro-F1},
        symbolic x coords={
          Bitcoin-Alpha,
          Bitcoin-OTC,
          Slashdot,
          Epinions},
        xtick=data,
        xticklabel style={at={(0.5,-0.5)}, text width=1.5cm, align=center, font=\scriptsize, rotate=30}, %
    ]
    \addplot [draw=blue!60, fill=blue!20] coordinates {
      (Bitcoin-Alpha, 0.921)
      (Bitcoin-OTC, 0.901)
      (Slashdot, 0.891) 
      (Epinions, 0.920)};
    \addplot [draw=blue!100, fill=blue!60] coordinates {
      (Bitcoin-Alpha, 0.922)
      (Bitcoin-OTC, 0.892)
      (Slashdot, 0.875) 
      (Epinions, 0.831)};
  \end{axis}
\end{tikzpicture}
\begin{tikzpicture}
\footnotesize
  \centering
  \begin{axis}[
        ybar=0pt, 
        height=2.9cm, width=5.7cm,
        bar width=7.5pt,
        ymajorgrids=true, 
        tick align=inside,
        major grid style={line width=0.1pt,draw=white},
        enlarge x limits=0.2,
        ymin=0.68, ymax=0.92,
        axis x line*=left,
        axis y line*=left,
        ylabel={Macro-F1},
        symbolic x coords={
          Bitcoin-Alpha,
          Bitcoin-OTC,
          Slashdot,
          Epinions},
        xtick=data,
        xticklabel style={at={(0.5,-0.5)}, text width=1.5cm, align=center, font=\scriptsize, rotate=30}, %
    ]
    \addplot [draw=blue!60, fill=blue!20] coordinates {
      (Bitcoin-Alpha, 0.721)
      (Bitcoin-OTC, 0.773)
      (Slashdot, 0.765) 
      (Epinions, 0.902)};
    \addplot [draw=blue!100, fill=blue!60] coordinates {
      (Bitcoin-Alpha, 0.723)
      (Bitcoin-OTC, 0.751)
      (Slashdot, 0.704) 
      (Epinions, 0.787)};
  \end{axis}
\end{tikzpicture}
\caption{Sign prediction accuracies of \ours\ and \ours(All). 
} 
\label{fig:rq2-2_sampler}
\end{figure*}
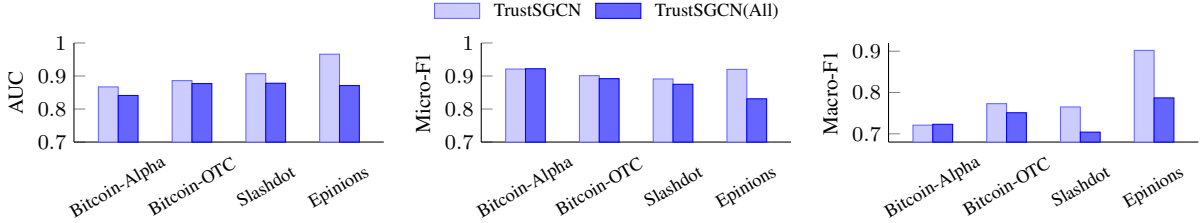

\vspace{1mm}
\noindent{\bf Results for EQ1.} 
To validate the effectiveness of our key ideas regarding trustworthiness-aware propagation, we conducted experiments to answer the following four sub-questions: 
\begin{itemize}[leftmargin=*]
 \item{\bf EQ1-1 (Trustworthiness Awareness)}: Does considering the trustworthiness for high-order signed relationships, inferred by balance theory, help to improve sign prediction accuracies?
 \item{\bf EQ1-2 (Random Sampling)}: Does the random sampling of each node's neighbors, used for embedding propagation, help to improve sign prediction accuracies?
 \item{\bf EQ1-3 (Propagation Ratios)}: Does incorporating the statistics of a given network for untrustworthy edge signs help to improve sign prediction accuracies?
 \item{\bf EQ1-4 (Path-length-based Attention)}: Does using learnable attention to control the propagation weight based on a path length help to improve sign prediction accuracies? 
\end{itemize}

For EQ1-1, 
we use two variants of \ours\ that do not employ \utgcn: (1) \ours(\tgcn), which performs embedding propagation based on the edge sings predicted by balance theory (\ie, using only \tgcn);
and (2) \ours(\fextra), which performs embedding propagation based on the edge signs predicted by \fextra, instead of balance theory. 
Table~\ref{table:eq2-1} demonstrates that \ours\ consistently outperforms both variants
across all datasets. 
We observed that the ratio of neighbors with untrustworthy edge signs, among all the neighbors in each node's \egonet, exceeds 45\% on all datasets. 
This finding suggests that the predictions made by balance theory or FExtra are often incorrect, and correcting these incorrect predictions is beneficial to sign predictions more accurate.

For EQ1-2, we use a variant of \ours, called \ours(All), which uses all of each node's neighbors (without random sampling) during embedding propagation.
Figure~\ref{fig:rq2-2_sampler} illustrates that \ours\ outperforms \ours(All) across all datasets in almost all cases. 
Notably, we observed that this sampling strategy is significantly effective for large datasets such as Slashdot and Epinions.
The results can be attributed to the mitigation of the over-fitting and over-smoothing problems achieved by random sampling, which is consistent with the findings reported in previous studies~\cite{hamilton2017inductive,kipf2016semi}.

\begin{table*}[!t]
\centering
\vspace{0.0cm}
\caption{Sign prediction accuracies of \ours, \ours(\texttt{Uniform}), \ours(\texttt{Reverse}), and \ours(\texttt{Mean})}
\vspace{-0.25cm}
\label{table:eq2-3}
\resizebox{\textwidth}{!} {
 \renewcommand{\arraystretch}{1.5}
\begin{tabular}{c|ccc|ccc|ccc|ccc}
\toprule
\textbf{Datasets} & \multicolumn{3}{c|}{\textbf{Bitcoin-Alpha}} & \multicolumn{3}{c|}{\textbf{Bitcoin-OTC}}  & \multicolumn{3}{c|}{\textbf{Slashdot}} & \multicolumn{3}{c}{\textbf{Epinions}} \\ 
\textbf{Metrics} & \textbf{Micro-F1} & \textbf{Macro-F1} & \textbf{AUC} & \textbf{Micro-F1} & \textbf{Macro-F1} & \textbf{AUC} & \textbf{Micro-F1} & \textbf{Macro-F1} & \textbf{AUC} & \textbf{Micro-F1} & \textbf{Macro-F1} & \textbf{AUC} \\ \midrule

\textbf{\ours} &  \bf{0.921} & \bf{0.721} & \bf{0.867} & \bf{0.901} & \bf{0.773} & {\bf 0.886} & \bf{0.891} & \bf{0.765} & \bf{0.907} & \bf{0.920} & \bf{0.902} & \bf{0.966}  \\ 

\textbf{\ours(\uniform)} & 0.918 & 0.710 & 0.860 & 0.896 & 0.762 & 0.873 & 0.884 & 0.753 & 0.900 & 0.917 & 0.898 & 0.964  \\ 
\textbf{\ours(\reverse)} & 0.912 & 0.700 & 0.841 & 0.894 & 0.752 & 0.881 & 0.887 & 0.750 & 0.904 & 0.915 & 0.895 & 0.964  \\ \midrule
\textbf{\ours(\mean)} & 0.919 & 0.716 & 0.855 & 0.899 & 0.771 & 0.882 & 0.883 & 0.734 & 0.901 & 0.914 & 0.894 & 0.964  \\

\bottomrule
\end{tabular}
}
\vspace{-0.35cm}
\end{table*}

For EQ1-3, we use two variants of \ours\ that employ different propagation ratios in \utgcn: (1) \ours(\uniform), which \textit{uniformly} propagates to both positive and negative embeddings with an equal ratio (\ie, 50:50); and (2) \ours(\reverse), which \textit{differently} propagates to positive and negative embeddings using the inverse of the ratios used in our \utgcn.
Table~\ref{table:eq2-3} illustrates that \ours\ consistently and significantly outperforms both variants.
The results demonstrate that leveraging the intrinsic properties of the input network is the most effective approach to avoid incorrect propagation when propagating to signed embeddings.

For EQ1-4, we use a variant of \ours, called \ours(\mean), which uses mean pooling instead of an attention mechanism during embedding propagation.
Table~\ref{table:eq2-3} indicates that using the attention mechanism yields higher sign prediction accuracy compared to mean pooling.
It demonstrates the significance of assigning different weights to nodes during embedding propagation based on their path lengths.

\begin{table}[t]
\footnotesize
\centering
\caption {The effect of $\lambda$ on sign prediction accuracies of \ours} 
\label{table:eq4-1}
\resizebox{0.49\textwidth}{!} {
 \renewcommand{\arraystretch}{1.8}
\begin{tabular}{cc|cccccc}
\toprule
\multicolumn{1}{c}{\multirow{1}{*}{\textbf{Datasets}}} & \multicolumn{1}{c|}{\multirow{1}{*}{\textbf{Metrics}}} & \multicolumn{1}{c}{\multirow{1}{*}{\boldsymbol{$\lambda$}\textbf{=0}}} &
\multicolumn{1}{c}{\multirow{1}{*}{\boldsymbol{$\lambda$}\textbf{=0.2}}} &
\multicolumn{1}{c}{\multirow{1}{*}{\boldsymbol{$\lambda$}\textbf{=0.4}}} &
\multicolumn{1}{c}{\multirow{1}{*}{\boldsymbol{$\lambda$}\textbf{=0.6}}} &
\multicolumn{1}{c}{\multirow{1}{*}{\boldsymbol{$\lambda$}\textbf{=0.8}}} &
\multicolumn{1}{c}{\multirow{1}{*}{\boldsymbol{$\lambda$}\textbf{=1}}} \\ \midrule

{\multirow{3}{*}{\textbf{Bitcoin-Alpha}}} & \textbf{Micro-F1} &  0.917 & 0.912 & 0.917 & 0.920 & 0.916 & {\bf 0.921}  \\ 
 & \textbf{Macro-F1}& 0.705 & 0.698 & 0.705 & {\bf 0.723} & 0.700 & 0.721  \\ 
 &  \textbf{AUC} & 0.861 & 0.859 & 0.861 & 0.865 & 0.861 & \bf {0.867} \\ \midrule

{\multirow{3}{*}{\textbf{Bitcoin-OTC}}}& \textbf{Micro-F1} &  0.895 & 0.899 & 0.895 & 0.896 & \bf{0.901} & 0.898  \\ 
 & \textbf{Macro-F1}& 0.762 & 0.772 & 0.762 & 0.762 & \bf{0.773} & 0.770  \\ 
 & \textbf{AUC} & 0.881 & 0.883 & 0.881 & 0.882 & \bf{0.886} & 0.881 \\ \midrule

{\multirow{3}{*}{\textbf{Slashdot}}} & \textbf{Micro-F1} &  0.889 & 0.890 & 0.885 & 0.881 & 0.888 & \bf{0.891}  \\ 
 & \textbf{Macro-F1}& 0.750 & 0.756 & 0.751 & 0.743 & 0.750 & \bf{0.765}  \\ 
 & \textbf{AUC} & 0.904 & 0.905 & 0.904 & 0.901 & 0.905 & \bf{0.907} \\ \midrule
 
{\multirow{3}{*}{\textbf{Epinions}}} & \textbf{Micro-F1} &  0.910 & 0.911 & 0.911 & 0.912 & 0.911 & \bf{0.919}  \\ 
 & \textbf{Macro-F1}& 0.890 & 0.891 & 0.890 & 0.892 & 0.890 & \bf{0.901}  \\ 
 & \textbf{AUC} & 0.960 & 0.961 & 0.962 & 0.963 & 0.964 & \bf{0.966} \\ \bottomrule
\end{tabular}
}
\vspace{-0.35cm}
\end{table}

\vspace{1mm}
\noindent{\bf Results for EQ2.} 
To evaluate the effectiveness of the joint learning of $\mathcal{L}_{sign}$ and $\mathcal{L}_{status}$, we investigate the impact of the weight parameter $\lambda$ for the status loss on the sign prediction performance of \ours. 
Table~\ref{table:eq4-1} presents the results.
It demonstrates that the accuracy of \ours\ is the lowest when $\lambda$=0, while it is the highest when $\lambda$=1 in almost all cases. 
The results emphasize the importance of simultaneously learning both the sign and direction information of the input network through $L_{sign}$ and $L_{status}$.

\begin{table*}[!t]
\footnotesize
\centering
\vspace{0.0cm}
\caption{Sign prediction accuracies of 5 competitors and \ours\ in terms of AUC}
\vspace{-0.25cm}
\label{table:comparison}
\resizebox{\textwidth}{!} {
 \renewcommand{\arraystretch}{1.3}
\begin{tabular}{c|cccccc|cccccc}
\toprule
\textbf{Training Ratios} & \multicolumn{6}{c|}{\boldsymbol{$x$}\textbf{=80}} & \multicolumn{6}{c}{\boldsymbol{$x$}\textbf{=60}}  \\ 
\textbf{Methods}  & \textbf{SGCN} & \textbf{SiGAT} & \textbf{SNEA} & \textbf{SDGNN} & \textbf{SGCL} & \textbf{\ours} & \textbf{SGCN} & \textbf{SIGAT} & \textbf{SNEA} & \textbf{SDGNN} & \textbf{SGCL} & \textbf{\ours} \\ \midrule
\textbf{Bitcoin-Alpha} &  0.689 & 0.837 & 0.805 & 0.838 & \ul{0.840} & {\bf 0.867} & 0.702 & 0.825 & 0.775 & \ul{0.835} & 0.832 & \bf{0.851}  \\ 
\textbf{Bitcoin-OTC} & 0.763 & 0.875 & 0.805 & 0.876 & \ul{0.882} & {\bf 0.886} & 0.740 & 0.871 & 0.796 & 0.870 & \ul{0.873} & \bf{0.887}  \\ 
\textbf{Slashdot} & 0.805 & \ul{0.894} & 0.788 & 0.876 & 0.858 & {\bf 0.907} & 0.791 & \ul{0.892} & 0.791 & 0.871 & 0.843 & \bf{0.907}  \\ 
\textbf{Epinions} & 0.920 & \ul{0.961} & 0.850 & \ul{0.961} & 0.947 & {\bf 0.966} & 0.919 & \ul{0.954} & 0.837 & 0.950 & 0.951 & \bf{0.961}  \\  \midrule
\textbf{Training Ratios}& \multicolumn{6}{c|}{\boldsymbol{$x$}\textbf{=40}} & \multicolumn{6}{c}{\boldsymbol{$x$}\textbf{=20}}  \\ 
\textbf{Methods}   & \textbf{SGCN} & \textbf{SiGAT} & \textbf{SNEA} & \textbf{SDGNN} & \textbf{SGCL} & \textbf{\ours} & \textbf{SGCN} & \textbf{SIGAT} & \textbf{SNEA} & \textbf{SDGNN} & \textbf{SGCL} & \textbf{\ours} \\ \midrule
\textbf{Bitcoin-Alpha} &  0.681 & 0.809 & 0.771 & 0.826 & \ul{0.827} & {\bf 0.833} & 0.677 & 0.764 & 0.700 & 0.778 & \ul{0.781} & \bf{0.788}  \\ 
\textbf{Bitcoin-OTC} & 0.732 & \ul{0.858} & 0.804 & 0.854 & 0.852 & {\bf 0.875} & 0.672 & 0.819 & 0.746 & \ul{0.825} & 0.798 & \bf{0.843}  \\ 
\textbf{Slashdot} & 0.773 & \ul{0.878} & 0.779 & 0.870 & 0.852 & {\bf 0.895} & 0.777 & \ul{0.860} & 0.763 & 0.853 & 0.837 & \bf{0.878}  \\ 
\textbf{Epinions} & 0.921 & \ul{0.948} & 0.818 & \ul{0.948} & 0.937 & {\bf 0.956} & 0.906 & 0.927 & 0.794 & \ul{0.940} & 0.922 & \bf{0.942}  \\  
\bottomrule
\end{tabular}
}
\vspace{-0.1cm}
\end{table*}

\vspace{1mm}
\noindent{\bf Results for EQ3.} 
We conducted comparative experiments on a sign prediction task to demonstrate the superiority of \ours\ over 5 state-of-the-art SNE methods.
To do this, we use $x$(=80, 60, 40, 20)\% of the existent edges in each dataset as the training set and the remaining edges as the test set.
That is, as the value of $x$ decreases, the sparsity of the training set increases.
Table~\ref{table:comparison} shows the results for AUC. 
The values boldfaced and \ul{underlined} indicate the best and 2nd-best accuracies in each row (\ie, each dataset), respectively.

We summarize the results shown in Table~\ref{table:comparison} as follows.
First, among the competitors, no single method consistently outperforms the others.
That is, the best competitors vary per dataset and training ratio. 
Second and most importantly, \ours\ universally outperforms \emph{all} competitors in \emph{all} settings on \emph{all} datasets. 
Notably, we highlight that as the value of $x$ decreases, the accuracy improvement of \ours\ over competitors tends to increase.
Specifically, on Bitcoin-OTC, \ours\ yields up to 0.45\% and 2.18\% higher AUC than the best competitor (\ie, SGCL and SDGNN) when $x$ is 80 and 20, respectively. Moreover, on Slashdot, \ours\ yields up to 1.45\% and 2.09\% higher AUC than the best competitor (\ie, SiGAT) when $x$ is 80 and 20, respectively. 
The results show that considering the trustworthiness on edge signs is helpful in accurately preserving the proximities between nodes (in particular, when information on nodes is insufficient), which leads to improving the accuracy of the sign prediction task.

\vspace{1mm}
\noindent{\bf Results for EQ4.} 
We analyze the change of the accuracy according to different values of $\beta$ (\ie, a threshold for $Trust(\hat{s}_{ij})$) in \ours.
Figure~\ref{fig:eq5} shows the results,
where the $x$-axis and $y$-axis represent the values of $\beta$ and AUCs, respectively. 
The optimal value of $\beta$ that yields the highest accuracy varies across different datasets. 
Specifically, for Bitcoin-Alpha, Bitcoin-OTC, Slashdot, and Epinions, the highest accuracy is achieved when $\beta$ is set to 0.80, 0.95, 1.00, and 1.00, respectively.
Overall, to achieve higher accuracy in the sign prediction task, it is crucial to set $\beta$ to a relatively higher value. 
This indicates that we should exploit the predicted edge signs only when the predictions by FExtra are sufficiently confident.

\begin{figure*}[t]
\footnotesize
\centering
\vspace{-0.08cm}
\begin{tikzpicture}
\begin{axis}[
title=\textbf{(a) Bitcoin-Alpha},
height=2.6cm, width=8.5cm,
xtick={1, 2, 3, 4, 5, 6, 7, 8, 9, 10,11},
xticklabels={0.2, 0.4, 0.6, 0.8, 0.9, 0.95, 0.96, 0.97, 0.98, 0.99, 1},
ylabel=AUC,
xlabel=$\beta$,
ymin=0.84, ymax=0.875,
y tick label style={/pgf/number format/.cd,fixed,fixed zerofill,precision=2,/tikz/.cd},]
\addplot[color=blue,mark=o,]
coordinates {(1, 0.853)(2,0.852)(3,0.854)(4,0.867)(5,0.866)(6,0.860)(7,0.858)(8,0.859)(9,0.860)(10,0.859)(11,0.861)};
\end{axis}
\end{tikzpicture}
\begin{tikzpicture}
\begin{axis}[
title=\textbf{(b) Bitcoin-OTC},
height=2.6cm, width=8.5cm,
xtick={1, 2, 3, 4, 5, 6, 7, 8, 9, 10,11},
xticklabels={0.2, 0.4, 0.6, 0.8, 0.9, 0.95, 0.96, 0.97, 0.98, 0.99, 1},
ylabel=AUC,
xlabel=$\beta$,
ymin=0.86, ymax=0.895,
y tick label style={/pgf/number format/.cd,fixed,fixed zerofill,precision=2,/tikz/.cd},]
\addplot[color=blue,mark=o,]
coordinates {(1, 0.875)(2,0.876)(3,0.872)(4,0.873)(5,0.874)(6,0.886)(7,0.883)(8,0.884)(9,0.884)(10,0.881)(11,0.883)};
\end{axis}
\end{tikzpicture}

\vspace{+0.2cm}
\begin{tikzpicture}
\begin{axis}[
title=\textbf{(c) Slashdot},
height=2.6cm, width=8.5cm,
xtick={1, 2, 3, 4, 5, 6, 7, 8, 9, 10,11},
xticklabels={0.2, 0.4, 0.6, 0.8, 0.9, 0.95, 0.96, 0.97, 0.98, 0.99, 1},
ylabel=AUC,
xlabel=$\beta$,
ymin=0.87, ymax=0.92,
y tick label style={/pgf/number format/.cd,fixed,fixed zerofill,precision=2,/tikz/.cd},]
\addplot[color=blue,mark=o,]
coordinates {(1, 0.888)(2,0.885)(3,0.885)(4,0.887)(5,0.889)(6,0.887)(7,0.889)(8,0.901)(9,0.904)(10,0.900)(11,0.907)};
\end{axis}
\end{tikzpicture}
\begin{tikzpicture}
\begin{axis}[
title=\textbf{(d) Epinions},
height=2.6cm, width=8.5cm,
xtick={1, 2, 3, 4, 5, 6, 7, 8, 9, 10,11},
xticklabels={0.2, 0.4, 0.6, 0.8, 0.9, 0.95, 0.96, 0.97, 0.98, 0.99, 1},
ylabel=AUC,
xlabel=$\beta$,
ymin=0.955, ymax=0.97,
y tick label style={/pgf/number format/.cd,fixed,fixed zerofill,precision=2,/tikz/.cd},]
\addplot[color=blue,mark=o,]
coordinates {(1, 0.960)(2,0.962)(3,0.962)(4,0.961)(5,0.961)(6,0.960)(7,0.961)(8,0.962)(9,0.963)(10,0.962)(11,0.966)};
\end{axis}
\end{tikzpicture}
\vspace{-0.25cm}
\caption{The effect of \boldsymbol{$\beta$} 
in a sign prediction task.}\label{fig:eq5}
\vspace{0.cm}
\end{figure*}
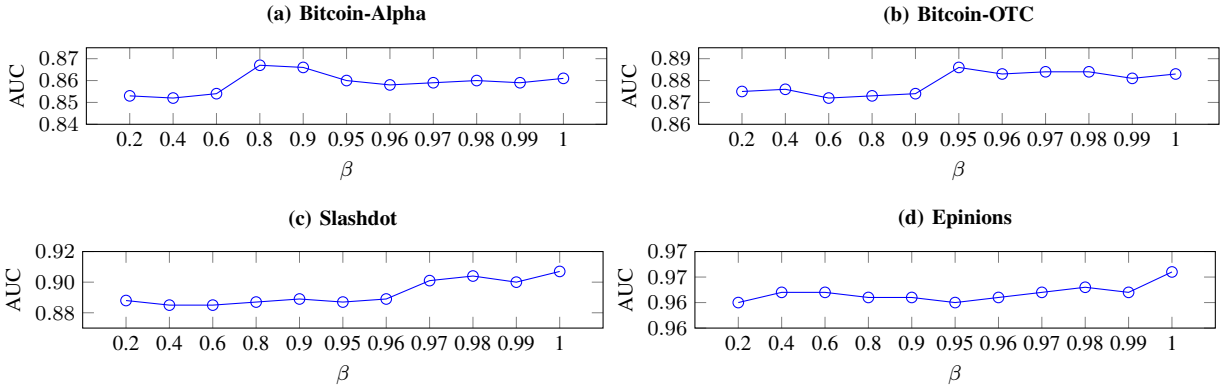
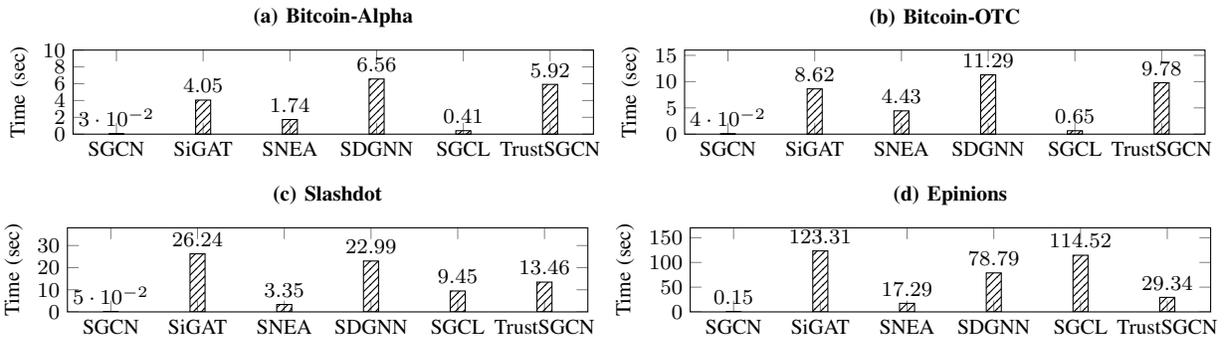
\begin{figure*}[t]
\footnotesize
\centering
\vspace{0.cm}
\begin{tikzpicture}
    \footnotesize
    \begin{axis}[
    title=\textbf{(a) Bitcoin-Alpha},
        width=8.5cm,
        height=2.7cm,
        bar width=0.2cm,
        ylabel={Time (sec)},
        ymin=0, ymax=10,
        xticklabels={SGCN, SiGAT, SNEA, SDGNN, SGCL, \ours},
        xtick={10, 20, 30, 40, 50, 60},
        ylabel near ticks,
        xlabel near ticks,
        nodes near coords,
        every node near coord/.append style={/pgf/number format/precision=2},
        ]
        \addplot[ybar,fill=black, pattern color = black, pattern=north east lines] coordinates {
        (10,0.03)(20,4.05)(30,1.74)(40,6.56)(50,0.41)(60,5.92)
        };
    \end{axis}
    \end{tikzpicture}    
    \begin{tikzpicture}
    \begin{axis}[
    title=\textbf{(b) Bitcoin-OTC},
        width=8.5cm,
        height=2.7cm,
        bar width=0.2cm,
        ylabel={Time (sec)},
        ymin=0, ymax=16,
        xticklabels={SGCN, SiGAT, SNEA, SDGNN, SGCL, \ours},
        xtick={10, 20, 30, 40, 50, 60},
        ylabel near ticks,
        xlabel near ticks,
        nodes near coords,
        every node near coord/.append style={/pgf/number format/precision=2},
        ]
        \addplot[ybar,fill=black, pattern color = black, pattern=north east lines] coordinates {
        (10,0.04)(20,8.62)(30,4.43)(40,11.29)(50,0.65)(60,9.78)
        };
    \end{axis}
    \end{tikzpicture}

    \vspace{+0.2cm}
    \begin{tikzpicture}
    \footnotesize
    \begin{axis}[
    title=\textbf{(c) Slashdot},
        width=8.5cm,
        height=2.7cm,
        bar width=0.2cm,
        ylabel={Time (sec)},
        ymin=0, ymax=38,
        xticklabels={SGCN, SiGAT, SNEA, SDGNN, SGCL, \ours},
        xtick={10, 20, 30, 40, 50, 60},
        ylabel near ticks,
        xlabel near ticks,
        nodes near coords,
        every node near coord/.append style={/pgf/number format/precision=2},
        ]
        \addplot[ybar,fill=black, pattern color = black, pattern=north east lines] coordinates {
        (10,0.05)(20,26.24)(30,3.35)(40,22.99)(50,9.45)(60,13.46)
        };
    \end{axis}
    \end{tikzpicture}
    \begin{tikzpicture}
    \footnotesize
    \begin{axis}[
    title=\textbf{(d) Epinions},
        width=8.5cm,
        height=2.7cm,
        bar width=0.2cm,
        ylabel={Time (sec)},
        ymin=0, ymax=170,
        xticklabels={SGCN, SiGAT, SNEA, SDGNN, SGCL, \ours},
        xtick={10, 20, 30, 40, 50, 60},
        ylabel near ticks,
        xlabel near ticks,
        nodes near coords,
        every node near coord/.append style={/pgf/number format/precision=2},
        ]
        \addplot[ybar,fill=black, pattern color = black, pattern=north east lines] coordinates {
        (10,0.15)(20,123.31)(30,17.29)(40,78.79)(50,114.52)(60,29.34)
        };
    \end{axis}
    \end{tikzpicture}
    \caption{Training times of 5 competitors and \ours.}\label{fig:eq6}
\vspace{-0.5cm}
\end{figure*}

\vspace{1mm}
\noindent{\bf Results for EQ5.} 
We conducted a comparison of the training times between \ours\ and its competitors.
Figure~\ref{fig:eq6} presents the results when we run a single epoch. 
The $x$-axis and the $y$-axis represent each method and the training time, respectively.
We observe that, while \ours\ may not be the fastest method, its training time is still acceptable, especially when we consider the fact that 
\ours\ significantly and consistently outperforms the competitors in terms of accuracy.
Therefore, we can conclude that \ours\ achieves higher accuracy than its competitors while requiring a reasonable training time.

\section{Conclusions and Future Work}\label{sec:conclusions}

In this work, we aimed to address the limitation of existing GCN-based SNE methods, which arise when relying blindly on the rules of balance theory.
To overcome the limitation, we proposed a novel SNE method, named \ours, which learns node embeddings based on the trustworthiness on edge signs for signed graph convolutional networks.
It consists of the generation of EgoNets, the measurement of trustworthiness on edge signs, and the trustworthiness-aware propagation of embeddings. 
Additionally, we incorporated both balance and status theories into a loss function so that the sign and direction of edges can be preserved in the embedding space.  

We conducted extensive experiments on four real-world signed network datasets to compare the performance of \ours\ with those of five existing GCN-based SNE methods. 
The results demonstrated that \ours\ consistently outperforms these methods in terms of sign prediction accuracy. 
We believe that our contribution is significant as we are the first to design a GCN architecture that explicitly considers the trustworthiness on edge signs.
We hope that our work will encourage follow-up studies on trustworthy SNE research~\cite{zhang2022trustworthy}. 

As future work, we plan to investigate an efficient way to reduce the computational time for preprocessing tasks such as FExtra-based sign prediction and calculation of triad statistics. 
Furthermore, extending the applicability of \ours\ to other graph structures, such as multi-relational networks \cite{feng2019marine} or dynamic signed networks \cite{dang2018link}, would be an intriguing direction.

\bibliographystyle{IEEEtran}
\bibliography{bibliography}
\label{sec:biography}
\vspace{-35pt}
\begin{IEEEbiography}
[{\includegraphics[width=1in,height=1in,clip,keepaspectratio]{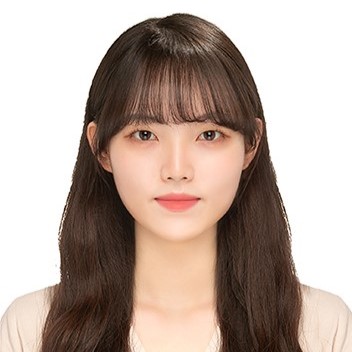}}]{Min-Jeong Kim} received the B.S. degree in Information and Communication Engineering from Chungbuk National University.
 She is currently a Ph.D. candidate in Artificial Intelligence at Hanyang University. Her research interests include graph mining, social network analysis, and recommender systems.

\end{IEEEbiography}

\vspace{-45pt}
\begin{IEEEbiography}[{\includegraphics[width=1in,height=1.25in,clip,keepaspectratio]{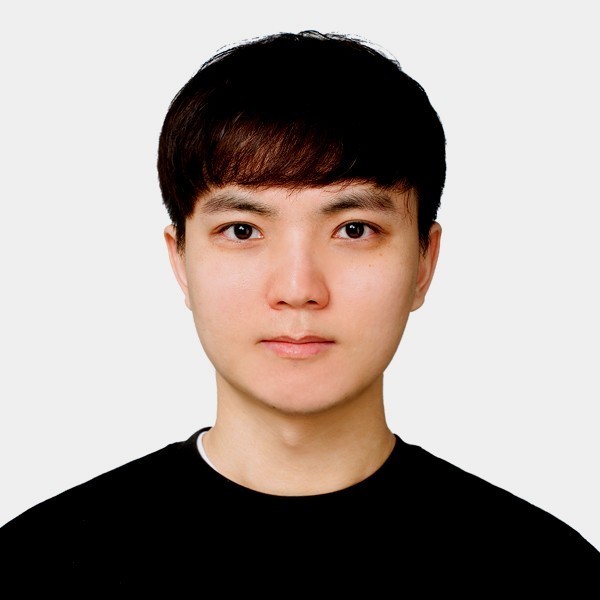}}]{Yeon-Chang Lee}
received the Ph.D. degree in computer science from Hanyang University in 2021 and the B.S. degree in medical IT from Eulji university in 2014. He is currently a postdoctoral researcher at Georgia Institute of Technology. His research interests include recommender systems, graph mining, and graph representation learning.
\end{IEEEbiography}

\vspace{-45pt}
\begin{IEEEbiography}[{\includegraphics[width=1in,height=1.25in,clip,keepaspectratio]{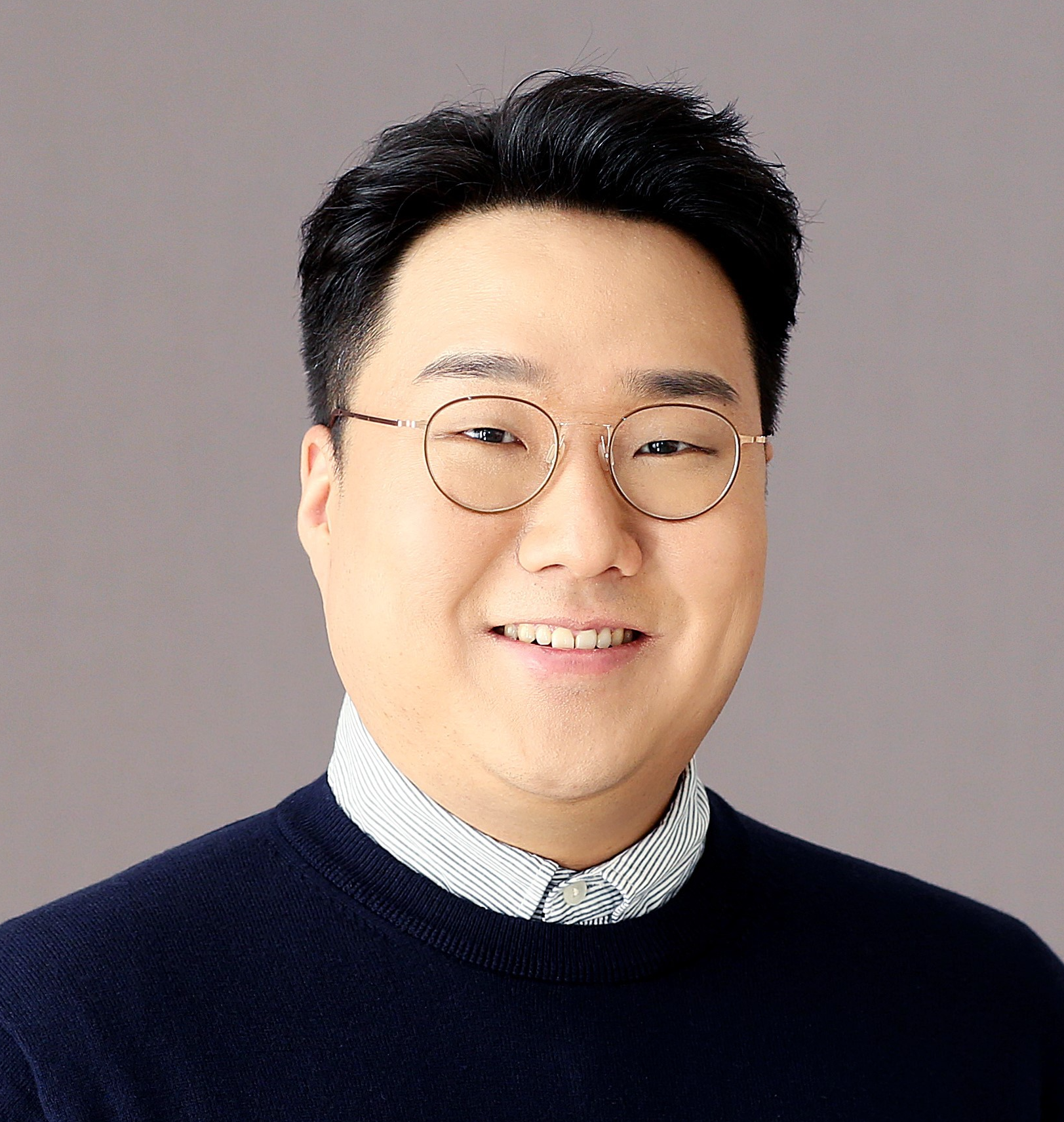}}]{David Y. Kang}
received the B.S. and Ph.D. degrees from Hanyang University in 2013 and 2022, respectively. He is currently a postdoctoral researcher at University of Michigan. His research interests include graph mining and social network analysis.
\end{IEEEbiography}

\vspace{-33pt}
\begin{IEEEbiography}[{\includegraphics[width=1in,height=1.25in,clip,keepaspectratio]{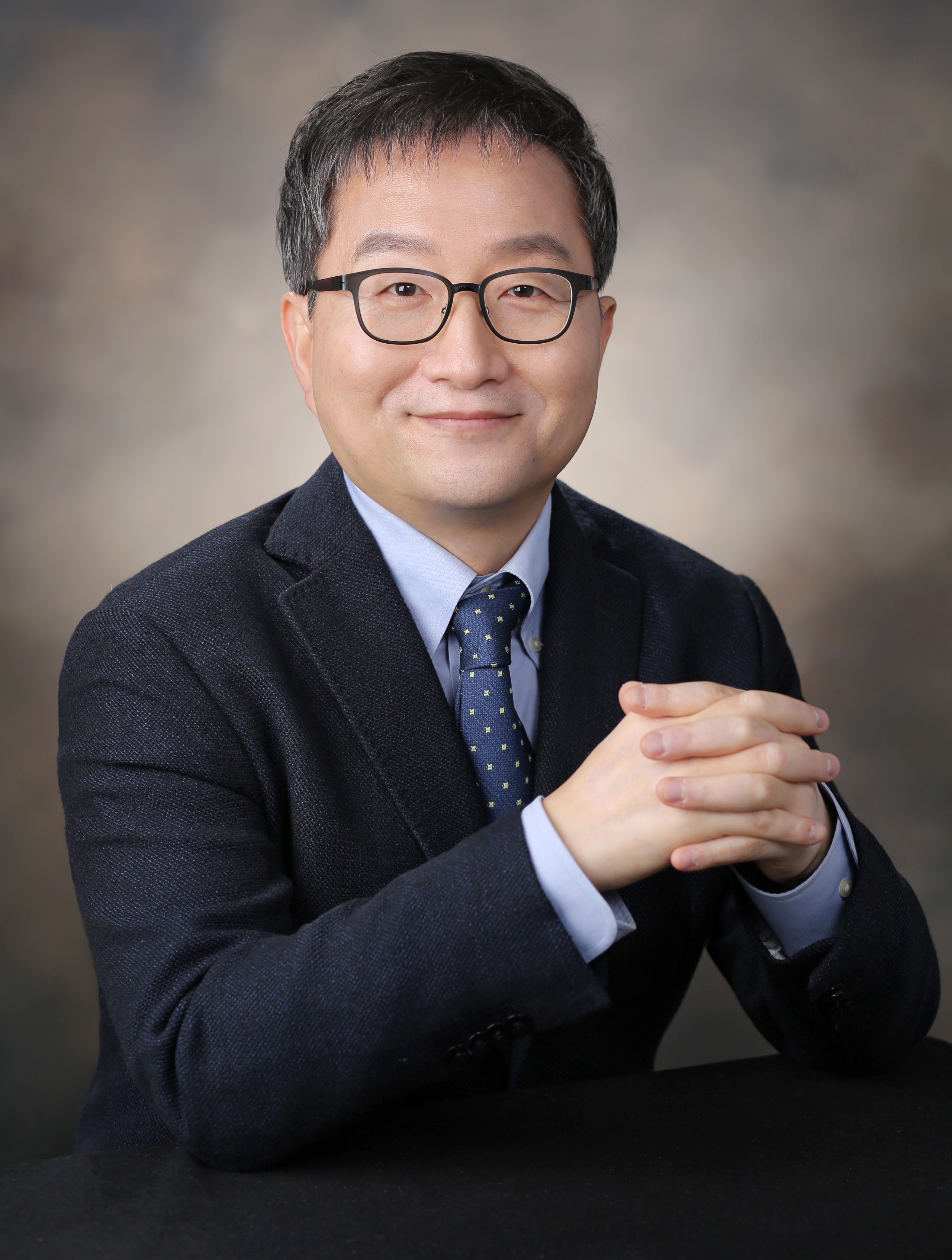}}]{Sang-Wook Kim}
received the B.S. degree in computer engineering from Seoul National University, in 1989, and the M.S. and Ph.D. degrees in computer science from the Korea Advanced Institute of Science and Technology (KAIST), in 1991 and 1994, respectively. From 1995 to 2003, he served as an associate professor with Kangwon National University. In 2003, he joined Hanyang University, Seoul, Korea, where he currently is a professor in the Department of Computer Science and the director of the Brain-Korea-21-FOUR research program. He is also leading a National Research Lab (NRL) Project funded by the National Research Foundation since 2015. From 2009 to 2010, he visited the Computer Science Department, Carnegie Mellon University, as a visiting professor. From 1999 to 2000, he worked with the IBM T. J. Watson Research Center, USA, as a postdoc. He also visited the Computer Science Department of Stanford University as a visiting researcher in 1991. He is an author of
more than 200 papers in refereed international journals and international conference proceedings. His research interests include databases, data mining, multimedia information retrieval, social network analysis, recommendation, and web data analysis. He is a member of the ACM and the IEEE.
\end{IEEEbiography}

\appendices
\section{Detailed description of topological features}~\label{app:feature}

We employed FExtra in (M2) to measure the trustworthiness of the edge signs predicted by the balance theory. 
FExtra serves as an additional approach capable of predicting high-order edge signs by leveraging the topological information associated with the two nodes. 
To this end, FExtra utilizes 23 features for each node pair ($v_i$,$v_j$).
Out of these 23 features, 7 features are related to the degrees of $v_i$ and $v_j$ (\ie, (f1) $\sim$ (f7)), while the remaining 16 features are related to the triads consisting of $v_i$, $v_j$, and their common neighbors in $\mathcal{G}$ (\ie, (f8) $\sim$ (f23)):

\begin{itemize}
    \item (f1): the number of $v_i$'s positive edges
    \item (f2): the number of $v_j$'s positive edges
    \item (f3): the number of $v_i$’s negative edges
    \item (f4): the number of $v_j$’s negative edges
    \item (f5): the total degrees of $v_i$
    \item (f6): the total degrees of $v_j$
    \item (f7): the total number of common neighbors of $v_i$ and $v_j$
    \item (f8), (f9), (f10), and (f11): the numbers of triads when the edge signs between $v_i$ and each common neighbor $v_z$ and those between $v_j$ and $v_z$ are both positive
    \item (f12), (f13), (f14), and (f15): the numbers of triads when the edge signs between $v_i$ and each common neighbor $v_z$ and those between $v_j$ and $v_z$ are positive and negative, respectively
    \item (f16), (f17), (f18), and (f19): the numbers of triads when the edge signs between $v_i$ and each common neighbor $v_z$ and those between $v_j$ and $v_z$ are negative and positive, respectively
    \item (f20), (f21), (f22), and (f23): the numbers of triads when the edge signs between $v_i$ and each common neighbor $v_z$ and those between $v_j$ and $v_z$ are both negative
\end{itemize}

\begin{figure}[h]
\centering
\vspace{0.0cm}
\includegraphics[width=0.95\linewidth]{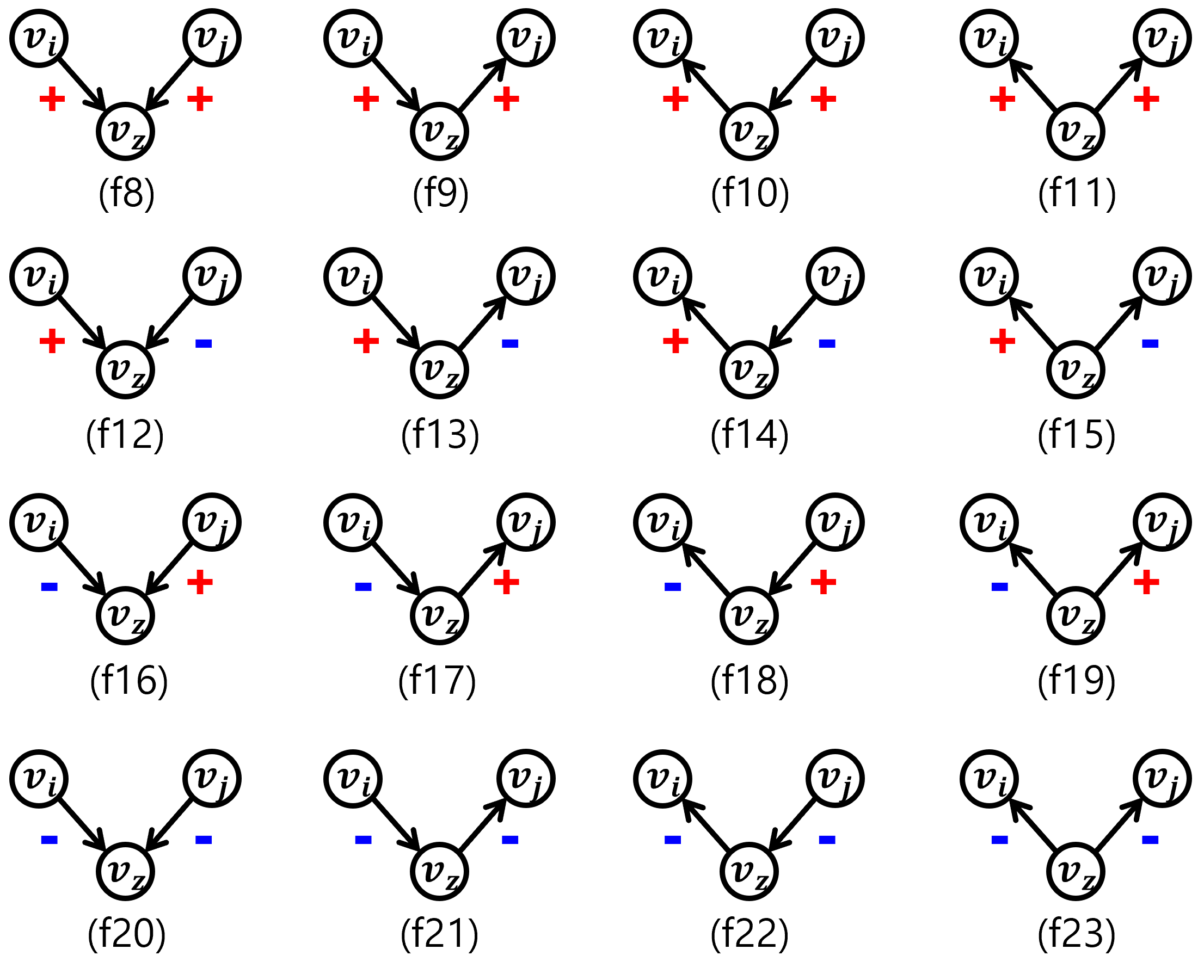}
\vspace{-0.3cm}
\caption{16 types of triads consisting of two nodes $v_i$, $v_j$ and a common neighbor $v_z$.}
\label{fig:23features}
\end{figure}

For features (f8) to (f23), the distinction is made based on whether the direction of the two edges is outgoing or incoming, as depicted in Figure~\ref{fig:23features}.
In other words, they are formed by the combination of four possible sign pairs and four possible direction pairs that the two edges can have.

\section{Further Results}~\label{app:results}    

\noindent\textbf{Further Results for EQ1-2.} In (M3) of \ours, we employ a sampling strategy to improve both training efficiency and accuracy. 
In addition to Figure~\ref{fig:rq2-2_sampler}, Table~\ref{table:gamma} shows the accuracy changes according to the number of randomly sampled nodes (\ie, $\gamma$).
We can see that using this sampling strategy is more effective than not using the strategy at all.

\vspace{1mm}
\noindent\textbf{Further Results for EQ4.} 
In (M2) of \ours, we verify whether $Trust(\hat{s}_{ij})$ exceeds the predefined threshold $\beta$, \ie, the second condition for determining the trustworthiness of edge signs. 
In addition to the AUC results (\ie, Figure~\ref{fig:eq5}), we also provide the results for other metrics in Table~\ref{table:beta}.
Similar to Figure~\ref{fig:eq5},  we can see that \ours\ is more accurate when the value of $\beta$ is relatively large.

\begin{table*}[t]
\footnotesize
\centering
\caption {The effect of $\gamma$ on sign prediction accuracies of \ours\ when we set $x$ to 80} 
\vspace{-0.25cm}
\label{table:gamma}
\resizebox{0.7\textwidth}{!} {
\renewcommand{\arraystretch}{1.2}
\begin{tabular}{cc|ccccccc}
\toprule
\multicolumn{1}{c}{\multirow{1}{*}{\textbf{Datasets}}} & 
\multicolumn{1}{c|}{\multirow{1}{*}{\textbf{Metrics}}} & 
\multicolumn{1}{c}{\multirow{1}{*}{\boldsymbol{$\gamma$}\textbf{=10}}} &
\multicolumn{1}{c}{\multirow{1}{*}{\boldsymbol{$\gamma$}\textbf{=20}}} &
\multicolumn{1}{c}{\multirow{1}{*}{\boldsymbol{$\gamma$}\textbf{=30}}} &
\multicolumn{1}{c}{\multirow{1}{*}{\boldsymbol{$\gamma$}\textbf{=40}}} &
\multicolumn{1}{c}{\multirow{1}{*}{\boldsymbol{$\gamma$}\textbf{=50}}} &
\multicolumn{1}{c}{\multirow{1}{*}{\boldsymbol{$\gamma$}\textbf{=60}}} &
\multicolumn{1}{c}{\multirow{1}{*}{\textbf{All}}}
\\ \midrule

{\multirow{3}{*}{\textbf{Bitcoin-Alpha}}} & \textbf{Micro-F1} &   0.914 & 0.917 &  \bf {0.921} & 0.916 & 0.918 & 0.917 & 0.919 \\ 
 & \textbf{Macro-F1}& 0.706 & 0.709 &  \bf {0.721} & 0.708 & 0.712 & 0.707 & 0.707 \\ 
 &  \textbf{AUC} & 0.851 & 0.860 & \bf {0.867} & 0.860 & 0.858 & 0.858 & 0.844 \\ \midrule

{\multirow{3}{*}{\textbf{Bitcoin-OTC}}}& \textbf{Micro-F1} &   0.897 & 0.897 & \bf {0.901} & 0.897 & 0.895 & 0.899 & 0.892 \\ 
 & \textbf{Macro-F1}& 0.767 & 0.767 & \bf {0.773} & 0.764 & 0.762 & 0.769 & 0.751\\ 
 & \textbf{AUC} & 0.884 & 0.883 & \bf {0.886} & 0.880 & 0.880 & 0.880 & 0.877 \\ \midrule

{\multirow{3}{*}{\textbf{Slashdot}}} & \textbf{Micro-F1} & 0.888 & \bf{0.891} & 0.887 & 0.885 & 0.890 & 0.887 & 0.880  \\ 
 & \textbf{Macro-F1}& 0.757 & \bf {0.765} & 0.759 & 0.758 & 0.757 & 0.755 & 0.728 \\ 
 & \textbf{AUC} & 0.906 & \bf {0.907} & 0.900 & 0.899 & 0.903  & 0.902 & 0.881 \\ \midrule
 
{\multirow{3}{*}{\textbf{Epinions}}} & \textbf{Micro-F1} &  \bf {0.920} & 0.919 & 0.918 & 0.917 & 0.915 & 0.914 & 0.901   \\ 
 & \textbf{Macro-F1}& \bf {0.902} & 0.901 & 0.900 & 0.898 & 0.896 & 0.895 & 0.880  \\ 
 & \textbf{AUC} & \bf {0.966} & 0.965 & 0.964 & 0.964 & 0.964 & 0.963 & 0.951  \\ \bottomrule
 
\end{tabular} 
}
\vspace{-0.35cm}

\end{table*}

\begin{table*} [hbt!]
\footnotesize
\centering
\caption {The effect of $\beta$ on sign prediction accuracies of \ours\ when we set $x$ to 80} 
\vspace{-0.25cm}
\label{table:beta}
\resizebox{\textwidth}{!} {
 \renewcommand{\arraystretch}{1.5}
\begin{tabular}{cc|ccccccccccc}
\toprule
\multicolumn{1}{c}{\multirow{1}{*}{\textbf{Datasets}}} & 
\multicolumn{1}{c|}{\multirow{1}{*}{\textbf{Metrics}}} & 
\multicolumn{1}{c}{\multirow{1}{*}{\boldsymbol{$\beta$}\textbf{=0.2}}} &
\multicolumn{1}{c}{\multirow{1}{*}{\boldsymbol{$\beta$}\textbf{=0.4}}} &
\multicolumn{1}{c}{\multirow{1}{*}{\boldsymbol{$\beta$}\textbf{=0.6}}} &
\multicolumn{1}{c}{\multirow{1}{*}{\boldsymbol{$\beta$}\textbf{=0.8}}} &
\multicolumn{1}{c}{\multirow{1}{*}{\boldsymbol{$\beta$}\textbf{=0.9}}} &
\multicolumn{1}{c}{\multirow{1}{*}{\boldsymbol{$\beta$}\textbf{=0.95}}} &
\multicolumn{1}{c}{\multirow{1}{*}{\boldsymbol{$\beta$}\textbf{=0.96}}} &
\multicolumn{1}{c}{\multirow{1}{*}{\boldsymbol{$\beta$}\textbf{=0.97}}} &
\multicolumn{1}{c}{\multirow{1}{*}{\boldsymbol{$\beta$}\textbf{=0.98}}} &
\multicolumn{1}{c}{\multirow{1}{*}{\boldsymbol{$\beta$}\textbf{=0.99}}} &
\multicolumn{1}{c}{\multirow{1}{*}{\boldsymbol{$\beta$}\textbf{=1}}}
\\ \midrule

{\multirow{3}{*}{\textbf{Bitcoin-Alpha}}} & \textbf{Micro-F1} &  0.913 & 0.917 & 0.914 & \bf {0.921} & 0.919 & 0.917 & 0.917 & 0.917 & 0.913 & 0.914 & 0.919 \\ 
 & \textbf{Macro-F1}& 0.689 & 0.698 & 0.695 & \bf {0.721} & 0.712 & 0.704 & 0.717 & 0.711 & 0.689 & 0.710 & 0.705  \\ 
 &  \textbf{AUC} & 0.853 & 0.852 & 0.854 & \bf {0.867} & 0.866 & 0.860 & 0.858 & 0.859 & 0.860 & 0.859 & 0.861 \\ \midrule

{\multirow{3}{*}{\textbf{Bitcoin-OTC}}}& \textbf{Micro-F1} &  0.890 & 0.892 & 0.891 & 0.894 & 0.895 & \bf {0.901} & 0.893 & 0.891 & 0.892 & 0.899 & 0.897 \\ 
 & \textbf{Macro-F1}& 0.745 & 0.748 & 0.751 & 0.751 & 0.758 & \bf {0.773} & 0.751 & 0.747 & 0.748 & 0.768 & 0.756\\ 
 & \textbf{AUC} & 0.875 & 0.876 & 0.872 & 0.873 & 0.874 & \bf {0.886} & 0.883 & 0.884 & 0.884 & 0.881 & 0.883 \\ \midrule

{\multirow{3}{*}{\textbf{Slashdot}}} & \textbf{Micro-F1} &  0.877 & 0.876 & 0.880 & 0.869 & 0.868 & 0.875 & 0.870 & 0.884 & 0.886 & 0.881 & \bf {0.891} \\ 
 & \textbf{Macro-F1}& 0.721 & 0.713 & 0.719 & 0.689 & 0.683 & 0.700 & 0.682 & 0.734 & 0.736 & 0.727 & \bf {0.765}  \\ 
 & \textbf{AUC} & 0.888 & 0.885 & 0.885 & 0.887 & 0.889 & 0.887 & 0.889  & 0.901 & 0.904 & 0.900 & \bf {0.907}\\ \midrule
 
{\multirow{3}{*}{\textbf{Epinions}}} & \textbf{Micro-F1} &  0.911 & 0.914 & 0.915 & 0.911 & 0.914 & 0.906 & 0.911 & 0.913 & 0.915 & 0.914 & \bf {0.920}  \\ 
 & \textbf{Macro-F1}& 0.891 & 0.894 & 0.896 & 0.891 & 0.894 & 0.882 & 0.891 & 0.893 & 0.896 & 0.895 & \bf {0.902}  \\ 
 & \textbf{AUC} & 0.960 & 0.962 & 0.962 & 0.961 & 0.961 & 0.960 & 0.961 & 0.962 & 0.963 & 0.962 & \bf {0.966} \\ \bottomrule
\end{tabular}
}
\vspace{-0.1cm}
\end{table*}

\end{document}